%% file: main.tex
\documentclass[twocolumn,showpacs,preprintnumbers,amsmath,amssymb,aps]{revtex4-1}

\usepackage{graphicx}
\usepackage{eurosym}
\usepackage[T1]{fontenc}
\usepackage[usenames,dvipsnames,table]{xcolor}
\usepackage[utf8]{inputenc}
\usepackage{lmodern}

\include{definitions}

\begin{document}

\newcommand{\kt}[1]{\textcolor{red}{[KT: #1]}}
\newcommand{\ymt}[1]{\textcolor{blue}{[YMT: #1]}}
\newcommand{\DP}[1]{{\textcolor{magenta}{(DP: #1)}}}

\newcommand{\raa}{R_{\rm AA}}

\newcommand{\rmd}{{\rm d}}


\title{Jet Suppression and Azimuthal Anisotropy from RHIC to LHC}

\author{Yacine Mehtar-Tani} 
\email{mehtartani@bnl.gov}
\affiliation{Physics Department, Brookhaven National Laboratory, Upton, NY 11973, USA}
\author{Daniel Pablos}
\email{daniel.pablos@usc.es}
\affiliation{Instituto Galego de F\'isica de Altas Enerx\'ias IGFAE, Universidade de Santiago de Compostela, E-15782 Galicia-Spain}
\affiliation{Departamento de F\'isica, Universidad de Oviedo, Avda. Federico Garc\'ia Lorca 18, 33007 Oviedo, Spain}
\affiliation{Instituto Universitario de Ciencias y Tecnolog\'ias Espaciales de Asturias (ICTEA), Calle de la Independencia 13, 33004 Oviedo, Spain}
\author{Konrad Tywoniuk}
\email{konrad.tywoniuk@uib.no}
\affiliation{Department of Physics and Technology, University of Bergen, Allégaten 55, 5007 Bergen, Norway}%

\date{\today}

\begin{abstract}
Azimuthal anisotropies of high-$p_T$ particles produced in heavy-ion collisions are understood as an effect of a geometrical selection bias. Particles oriented in the direction in which the QCD medium formed in these collisions is shorter, suffer less energy loss, and thus, are over-represented in the final ensemble  compared to those oriented in the direction in which the medium is longer. In this work we present the first semi-analytical predictions, including propagation through a realistic, hydrodynamical background, of the elliptic azimuthal anisotropy for jets, obtaining a quantitative agreement with available experimental data as function of the jet $p_T$, its cone size $R$ and the collisions centrality. Jets are multi-partonic, extended objects and their energy loss is sensitive to substructure fluctuations. This sensitivity is determined by the physics of color coherence that relates to the ability of the medium to resolve those partonic fluctuations. Specifically, color dipoles with an angular separation smaller than a critical angle, $\theta_c$, are not resolved by the medium and they effectively act as a coherent source of energy loss. We find that elliptic jet azimuthal anisotropy has a specially strong dependence on coherence physics due to the marked length-dependence of $\theta_c$. By combining our predictions for the collision systems and center of mass energies studied at RHIC and the LHC, covering a wide range of typical values of $\theta_c$, we show that the relative size of elliptic jet azimuthal anisotropies for jets with different cone-sizes $R$ follows a universal trend that indicates a transition from a coherent regime of jet quenching to a decoherent regime. These results suggest a way forward to reveal the role played by the physics of jet color decoherence in probing deconfined QCD matter.

\end{abstract}

\maketitle

\section{Introduction}
Ultrarelativistic heavy-ion collisions taking place at present-day colliders provide insights into the dynamics of deconfined nuclear matter created in the aftermath of these collisions~\cite{Busza:2018rrf}.  The remarkable correlation between the geometry of the initial collision state and the final state momentum anisotropies of the observed particles has inspired the conjecture that this new state of matter, referred to as the quark-gluon plasma (QGP), behaves like an expanding droplet of liquid deconfined nuclear matter \cite{Gale:2013da,Heinz:2013th}.
The surprising fact that this apparent hydrodynamic behavior is also observed in proton-nucleus and proton-proton collisions~\cite{Noronha:2024dtq}
has posed further challenges to our understanding of the applicability of hydrodynamics for smaller colliding systems, where spatial gradients are expected to be large. These challenges were put into a new perspective with the finding of hydrodynamic attractor solutions, present in a variety of microscopic descriptions, on which a longitudinally-boosted system will fall provided that the non-hydrodynamic modes of the theory decay sufficiently fast~\cite{Heller:2015dha,Romatschke:2017vte,Kurkela:2019set} (for a recent review see~\cite{Soloviev:2021lhs}). While these developments support an  effective description of deconfined Quantum Chromodynamics (QCD) matter using relativistic hydrodynamics, the details of the microscopic description of the system remain to be fully elucidated.

High-energy jets are rare events that are produced alongside the QGP with which they strongly interact while they travel through, offering  valuable information about the properties of the medium. Due to their multi-scale nature, spanning from the jet transverse momentum $\sim \pT$ to the non-perturbative scales of the vacuum $\sim \Lambda_\text{\tiny QCD}$, they probe the QGP at different length-scales. Through the study of the modifications imprinted in this in-medium developing multi-partonic system, we can infer the nature of the interaction between the energetic partons and the QGP. Moreover, jets that undergo significant modification or quenching will experience the absorption of a sizable portion of their energy and momentum by the medium. This occurs via dissipative processes, the study of which provides insights into the approach to thermal equilibrium as well as broader aspects of non-equilibrium QCD dynamics.  

A direct connection between jet modifications and the fundamental interactions between the QGP and fast partons crucially relies on our understanding of jet fragmentation in these systems. From the point of view of a high-energy jet, the modest energy scales that characterize a droplet of QGP are typically small compared to the largest jet energy scales. This implies that medium modifications set in at intermediate length-scales, preceded by a nearly vacuum-like evolution~\cite{Mehtar-Tani:2017web,Caucal:2018dla}. Nevertheless, as confirmed in experimental data, the impact of medium modifications is striking~\cite{Apolinario:2022vzg}. Jet yield suppression, typically quantified by the nuclear modification factor $R_{\rm AA}$, stands as a paradigmatic illustration of jet quenching. This phenomenon is interpreted as arising from the interaction between energetic colored objects and deconfined QCD matter. 

A complete description of the dynamics of deconfined QCD matter will however require establishing a consistent picture between the microscopic physics of the bulk of the system and that of its interaction with the jets that traverse it. Recent theoretical progress on our understanding of the intricate interplay between the jet and the medium scales is now allowing for first-principles phenomenological applications which can be meaningfully confronted to the high-precision data expected from upcoming runs at RHIC and LHC.

In a previous publication~\cite{Mehtar-Tani:2021fud} we provided the first analytical description of jet yield suppression as a function of the jet size $R$. In short, this description accounts both for the elastic and radiative processes induced by the interaction of a fundamental colored object, a parton, with the medium, as well as the process of resolving which partons participate in the medium interactions during the jet fragmentation. Our analytical calculations were convolved with a realistic model for the hydrodynamical background and compared to experimental data at LHC energies. We performed a thorough study of the relative sizes of the different sources of theoretical uncertainties and established that these are dominated by the computation of the resolved phase-space of the collinear in-vacuum parton cascade up to relatively large angles $R\sim 0.6$, which is systematically improvable within perturbative QCD (pQCD).
In the present work, we extend this framework to the study of jet suppression at RHIC energies. This extension serves to further validate the fundamental working assumptions based on factorizing jet production, fragmentation and subsequent medium modification.

We also address another paradigmatic observable that is jet azimuthal anisotropy. Roughly speaking, while $R_{{AA}}$ is sensitive to the average path length traveled by the jet, this observable is sensitive to path-length differences in jet suppression due to the relative orientation in the transverse plane of the jet direction $\phi$ with respect to the event plane of the collision. 

Analogously to how azimuthal anisotropies are quantified for the particles belonging to the bulk of the collision, jet azimuthal anisotropies are defined via the $\pT$-dependent Fourier coefficients of their azimuthal distribution, the $v_n(\pT)$, as 
\begin{equation}
    \frac{\rmd N}{\rmd \phi \rmd \pT}=\frac{\rmd N/\rmd \pT}{2 \pi}\left[1+2\sum_n v_n(\pT)\cos(n(\phi-\Psi_n))\right] \, ,
\end{equation}
where $\rmd N/\rmd \pT$ is the multiplicity in the given $\pT$-bin and $\Psi_n$ defines the $n$'th order event plane.
The behavior of $\pT$-inclusive $v_n$, dominated by soft particle production from the bulk of the system, are understood in terms of the preferred directions determined by the initial geometry of the collision, which translate into momentum anisotropies due to the conversion of pressure gradients into momentum flow via the final state hydrodynamic evolution ~\cite{Ollitrault:1992bk}. In contrast, high-$\pT$ $v_n$ are caused by a selection bias effect that results into larger yields in the directions in which the high-$\pT$ object has had to traverse less amount of medium, i.e. in the directions with least quenching~\cite{Wang:2000fq,Gyulassy:2000gk,Shuryak:2001me}.

Due to the dominant elliptical, almond-like shape characteristic of the collision of two fairly round objects at finite impact parameter $b$, the largest $v_n$ is typically the elliptical flow coefficient, $v_2$. Nevertheless, fluctuations in the positions of the incoming nucleons inside the nuclei~\cite{Alver:2010gr}, and sub-nucleonic fluctuations in the color field configurations inside each of the nucleons at the moment of the collision~\cite{Schenke:2012wb}, will give rise to finite values for the higher harmonics $v_3$, $v_4$, and so on. For the case of quenching-induced flow coefficients, also fluctuations in the process of energy loss will contribute to the value of all flow harmonics~\cite{Noronha-Hostler:2016eow}. 

There is extensive recent literature on the study of flow coefficients for single energetic particles, or high-$\pT$ hadrons (see, e.g.,~\cite{Noronha-Hostler:2016eow,Andres:2016iys,Kumar:2019uvu,Zigic:2019sth,Park:2019sdn,Arleo:2022shs,Zigic:2022xks,Karmakar:2023ity}). However, much less work has been devoted to the study of the flow coefficients for jets, and most of what is present is obtained from Monte Carlo jet quenching models~\cite{Du:2021pqa,Barreto:2022ulg,He:2022evt}. Given that jets are extended objects, consisting of a collection of hadrons within a given cone $R$, they are affected by a number of physical mechanisms that are not present in the description of single-hadron observables. Our analytical theoretical framework consistently combines the most dominant of such mechanisms, rendering it appropriate to address the more challenging scenario involving jets.

Even within the medium, high-energy jets experience part of their evolution as if they were approximately in vacuum~\cite{Caucal:2018dla}. This is due to the presence of a wide scale separation between the typical time of high-virtuality parton splitting, as determined by the DGLAP evolution equations, and the typical time to produce a medium-induced emission via multiple soft scatterings, as determined by the BDMPS-Z equations~\cite{Kurkela:2014tla,Mehtar-Tani:2017web,Caucal:2018dla}. The DGLAP evolution resums large logarithms of the cone angle $R$, i.e. $\sim \log 1/R$, and is responsible for jet ``energy loss'' in vacuum from emissions off an initial hard parton with angles larger than $R$. Hence, the jet spectrum becomes a function of the cone angle $R$ and transverse momentum $\pT$ \cite{Dasgupta:2014yra,Dasgupta:2016bnd,Kang:2016mcy}. We incorporate this QCD evolution in the present work as well. 

Vacuum-like radiation taking place within the cone $R$ will no longer contribute to energy loss in vacuum at leading-logarithmic (LL) accuracy. However, in the presence of a medium they will determine the amount of color charges that can source out-of-cone radiation via interactions with that medium. The enhancement of multiple color charge energy loss by a Sudakov double logarithm is attributed to the distinction in energy loss between real and virtual contributions. Consequently, a critical aspect in the understanding of jet energy loss lies in determining the size of the vacuum-generated in-cone phase-space resolved by the medium, which we refer to as the ``quenched'' phase-space.

The first observations of the close relation between the size of the ``quenched'' phase-space and the amount of energy loss were obtained using phenomenological models. These were able to show that the origin of the jet core narrowing observed in data is due to a selection bias towards those jets that experienced a narrower fragmentation during its vacuum-like evolution~\cite{Rajagopal:2016uip,Casalderrey-Solana:2016jvj}. Even though each individual jet does get broadened in the process of energy loss, the final ensemble is biased towards those jets that lost the least energy, the narrower ones, due to the steepness of jet production spectra. This same phenomenon has been shown to account for a large part of the dijet asymmetry modification of PbPb compared to pp~\cite{Milhano:2015mng}, even in the absence of path-length differences between the leading and subleading jet. 

On the theory side, the role of hard radiation that occurs immediately after the hard partonic collision in sourcing further (medium-induced) emissions was elucidated in a series of papers analyzing the so-called {\it antenna setup} \cite{MehtarTani:2010ma,Casalderrey-Solana:2011ule,Mehtar-Tani:2011hma,Mehtar-Tani:2011lic,Mehtar-Tani:2012mfa}. These calculations revealed the existence of the so-called decoherence time, associated to the typical time needed by medium interactions before being able to resolve the individual colors of the created pair of partons. The case when this time is comparable to the medium length determines a critical angle $\theta_c$, corresponding to a minimal angle below which jet splittings act coherently as a single color charge and which profoundly affects the quenching of jets in the medium \cite{CasalderreySolana:2012ef}. For a homogeneous plasma, the critical angle is defined as
\beq 
\theta_c \sim \frac{1}{\sqrt{\hat q L^3}}\,,
\eeq
where $L$ is the medium length and $\hat q$ is the jet transport coefficient that describes Brownian motion in transverse momentum space.
This phenomenon, called ``color decoherence'', leads to a modified joint probability of losing energy off an initially color correlated pair of particles \cite{Mehtar-Tani:2017ypq}. On the level of the jet spectrum in heavy-ion collisions, accounting for early resolved vacuum-like emissions inside the medium and accounting for their energy loss leads to an enhanced suppression factor at high-$\pT$ where the additional ``Sudakov factor'' scales with the ``quenched'' phase-space \cite{Mehtar-Tani:2017web}.

The physics of color decoherence explains why single inclusive hadron $R_{\rm AA}$ is larger than jet $R_{\rm AA}$ \cite{Mehtar-Tani:2017web}, i.e. $R_{\rm AA}^\text{jet}(\pT) < R_{\rm AA}^\text{hadron}(\pT)$, and how this fact is intimately related to the high-$z$ enhancement of the measured jet fragmentation function modification~\cite{Casalderrey-Solana:2018wrw,Caucal:2020xad}. The first phenomenological implementation of the inclusion of finite resolution effects, or to which degree do two colored charges generated by vacuum-like evolution engage with the medium independently, in a Monte-Carlo jet quenching shower algorithm was done in~\cite{Hulcher:2017cpt}, and later in close analogy to the theoretical description described above in \cite{Caucal:2018dla}. The amount of selection bias towards narrower jets has been shown to be strongly affected by the size of the resolution parameter, leaving clear imprints on the distribution of the relatively hard, measurable splitting selected by the SoftDrop~\cite{Larkoski:2015lea} grooming procedure (see \cite{Caucal:2019uvr,Casalderrey-Solana:2019ubu}) or the Dynamical Grooming~\cite{Mehtar-Tani:2019rrk} procedure (see \cite{Caucal:2021cfb,Pablos:2022mrx} as well as~\cite{Cunqueiro:2023vxl} for a closely related substructure observable). Lately, renewed interest in the energy-energy correlators observable~\cite{Basham:1977iq,Basham:1978bw,Basham:1978zq} has arisen in the field of jet quenching~\cite{Andres:2022ovj,Andres:2023xwr,Yang:2023dwc,Barata:2023zqg,Barata:2023bhh} due, in part, to its potential sensitivity to emergent medium scales such as $\theta_c$. 

The present work shows how the $R$-dependence of jet $v_2$ closely relates to the existence of this medium resolution scale, providing additional tools with which to experimentally constrain these physics and build a coherent picture of jet energy loss across all observables.
With a full-fledged embedding of our theoretical framework into a realistic heavy-ion environment, we show that we can describe currently available jet $v_2$ data with a very good agreement. Further, by studying the $R$-dependence of this observable, we find an interesting departure from the one found for jet $R_{\rm AA}$, consisting in a clear and persistent ordering of jet $v_2$ as a function of the jet $R$ across kinematic and centrality selections. With simple analytical estimates we are able to pin-point the origin of this new phenomenon, which is rooted in the marked length-dependence of the medium angular resolution scale, $\theta_c$, combined with the fact that jet azimuthal anisotropy is a length-differential jet suppression observable. Moreover, we exploit centrality evolution to target different values of the typical decoherence angle $\theta_c$, and in this way establish an arrange of predictions which actually collapse into universal, analytically understandable curves, when presented in terms of the ratio of the two most relevant angular scales of the system, i.e. $\theta_c/R$.

The rest of the paper is organized as follows. In Section~\ref{sec:formalism} we present our theoretical formalism and explain how we embed it into a realistic heavy-ion environment. In Section~\ref{sec:results} we present the results for jet $R_{\rm AA}$ and jet $v_2$ at RHIC and LHC energies as a function of $R$ using our full semi-analytical framework. Next, in Section~\ref{sec:discussion}, we gain further insight on the phenomenological implications of the key ingredients of our computation by analytically studying a simplified scenario that allows a transparent qualitative interpretation of the full results of Section~\ref{sec:results}. Finally, in Section~\ref{sec:conc} we conclude and look ahead.

\section{Theoretical formalism}
\label{sec:formalism}

\begin{figure}
    \centering
    \includegraphics[width=0.9\columnwidth]{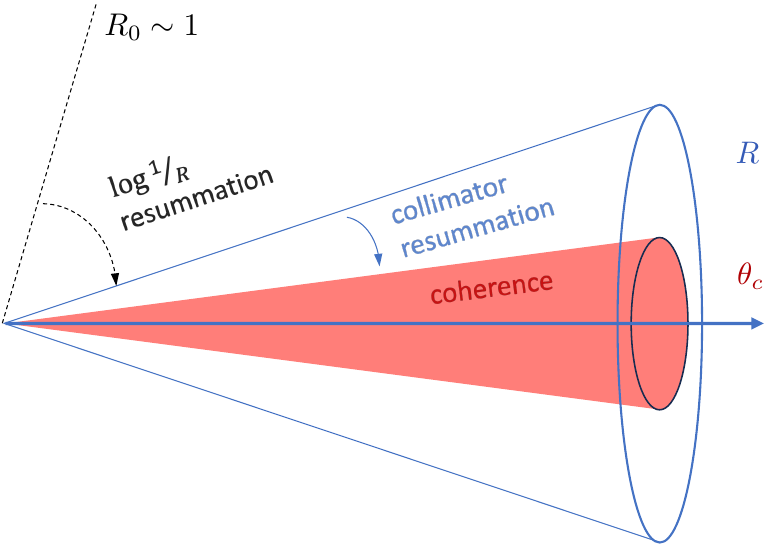}
    \caption{Sketch of the main steps involved in our computation. Starting from a large angular scale $R_0$, we evolve the jet spectrum down to a given cone $R$ using the microjets formalism. This evolution can lead to out-of-cone radiation due to vacuum physics. Depending on the amount of in-medium resolved charges between $R$ and the critical angle $\theta_c$, the collimator quantifies the shift in the jet spectrum for a given $R$ due to medium-induced effects.}
    \label{fig:sketch}
\end{figure}

In this section we provide a description of the theoretical formalism used to produce the full results of Section~\ref{sec:results}. These ingredients are essentially the same as those used to produce the results of publication~\cite{Mehtar-Tani:2021fud}, which we reproduce, often in a more detailed form, for convenience of the reader. The key elements are illustrated in Fig.~\ref{fig:sketch}.

\subsection{Vacuum-induced jet energy loss}

The cross-section to measure a jet with transverse momentum $\pT$ contained within a cone of extent $R$ in the $(\eta,\phi)$ angular plane can be computed in vacuum as~\cite{Dasgupta:2014yra} 
\beq
\label{eq:sigma-pp}
\sigma^{\tiny pp}(\pT,R) =  \sum_{k=q,g} f^{(n-1)}_{\jet/k} (R| \pT,R_0) \, \hat \sigma_k(\pT,R_0)\,,
\eeq
where $R_0$ is an initial large angular scale ($R_0 \sim 1$), and $n\equiv n_k(\pT,R_0)$ is the power-index of the cross-section of the hard parton with flavor $k$, $\hat \sigma_k(\pT,R_0)$. The latter is calculated at leading order (LO) at the factorization scale $Q^2_{\rm fac}$ via the convolution of the parton distribution functions (PDFs) $f_{i/p}(x,Q_\text{fac}^2)$ with the $2\to2$ QCD scattering cross section $\hat \sigma_{ij \to kl}$, i.e. $\hat \sigma_k = f_{i/p} \otimes f_{j/p} \otimes \hat \sigma_{ij \to kl} $. The moment of the fragmentation function of an initial hard parton with flavor $k$ is 
\beq
f^{(n)}_{\jet/k} (R| \pT,R_0) \equiv \int_0^1 \dd x \, x^{n} f_{\text{jet}/k}(R,x|\pT,R_0) \, ,
\eeq
receiving both quark and gluon contributions, i.e. $f^{(n)}_{\jet/k} = \sum_{i=q,g}f^{(n)}_{i/k}$, via flavor conversion during the DGLAP evolution \cite{Dasgupta:2014yra,Dasgupta:2016bnd,Kang:2016mcy,Dai:2016hzf}.
Here, $f_{\text{jet}/k}(R,x|\pT,R_0)$ is the probability of finding a sub-jet  carrying momentum fraction $x$ at cone-angle $R$ within a bigger jet with cone-angle $R_0$ and transverse momentum $\pT$.

The full, inclusive jet cross section  therefore becomes $\sigma^{\tiny pp}(\pT,R) = \sum_{i=q,g} \sigma_i^{\tiny pp}(\pT,R)$, where 
\beq
\label{eq:sigmai-pp}
\sigma_i^{\tiny pp}(\pT,R) =  \sum_{k=q,g} f^{(n-1)}_{i/k} (R| \pT,R_0) \, \hat \sigma_k(\pT,R_0) \,,
\eeq
is the spectrum for a final quark or gluon jet measured at cone angle $R$.

The DGLAP evolution is what describes the microjet spectrum at different angular scales $R$, effectively implementing out-of-cone radiation, or in other words, vacuum jet energy loss. Its evolution equations in terms of the moments of the fragmentation functions are
\begin{align}
\label{eq:lodglap}
\frac{d}{dt}f_q^{(n)}(t)&=\gamma^{(n)}_{qq}(t)f_q^{(n)}(t)+\gamma^{(n)}_{qg}(t)f_g^{(n)}(t) \nonumber \\
\frac{d}{dt}f_g^{(n)}(t)&=\gamma^{(n)}_{gg}(t)f_g^{(n)}(t)+\gamma^{(n)}_{gq}(t)f_q^{(n)}(t) \, 
\end{align}
where the evolution variable $t\equiv \ln\theta/R$, with $\theta>R$ and $R<R_0$, so that $t$ is evolved from $\ln R_0/R$ to 0. The anomalous dimensions at LO are \cite{Dasgupta:2014yra}
\begin{align}
    \gamma^{(n)}_{qq}(t)&=\int_0^1 \rmd z\, \frac{\alpha_s(t)}{\pi} p_{qq}(z)(z^n-1) \, ,\nonumber \\
    \gamma^{(n)}_{qg}(t)&=\int_0^1 \rmd z\,\frac{\alpha_s(t)}{\pi} p_{qg}(z)z^n \, ,\nonumber \\
    \gamma^{(n)}_{gg}(t)&=\int_0^1\rmd z\, \frac{\alpha_s(t)}{\pi} \big[p_{gg}(z)(z^n-z) - p_{gq}(z)\big] \, ,\nonumber \\
    \gamma^{(n)}_{gq}(t)&=\int_0^1 \rmd z\,\frac{\alpha_s(t)}{\pi} p_{gq}(z)z^n \, ,
    \end{align}
with the un-regularized Altarelli-Parisi (AP) splitting functions 
\begin{align}
    p_{qq}(z)&=C_F\frac{1+z^2}{1-z} \, ,\nonumber \\
    p_{gq}(z)&=C_F\frac{1+(1-z)^2}{z} \, ,\nonumber \\
    p_{gg}(z)&=C_A \frac{(1-z(1-z))^2}{z(1-z)} \, ,\nonumber \\
    p_{qg}(z)&=\frac{N_f}{2}(z^2+(1-z)^2) \, .
\end{align}
Finally, the running coupling $\alpha_s(t) = \alpha(k_t)$, with $k_t =z(1-z)p_T \theta$, is evaluated at LO with 5 active flavors and regularized as $\alpha_s(k_t) = \min[1,2\pi/(\beta_0 \log k_t/Q_0) ]$, with $\beta_0=23/3$ and $Q_0=0.09$ GeV.

In practice, we fit the shape of the initial spectra at $R=R_0=1$, for a given pseudorapidity $\eta$ range, using the event generator PYTHIA8 \cite{Sjostrand:2006za,Sjostrand:2007gs}, parameterized as $\dd \hat \sigma^{(k)}/\dd \pT = \sigma^{(k)}_0 \,(p^{(k)}_{{\sst T},0}/\pT)^{n^{(k)}(\pT)}$ and $n^{(k)}(\pT) = \sum_{i=0}^5 c^{(k)}_i \log^i (p^{(k)}_{{\sst T},0}/\pT )$. This effectively includes the LO $2\rightarrow2$ QCD scattering cross section convoluted with the PDFs (free-proton PDFs in pp collisions, nuclear PDFs in AA collisions), plus some large-angle evolution down to $R_0$.
We then use the LO DGLAP evolution equations~\eqref{eq:lodglap} to obtain the spectra for $R<1$ using Eq.~\eqref{eq:sigma-pp}, a step which we refer to as ``$\log 1/R$ resummation'' in the sketch depicted in Fig.~\ref{fig:sketch}. In pp collisions, and to LL accuracy, this completes the jet spectra computation. 

\subsection{Medium-induced jet energy loss}
\label{sec:medium-induced}
In the presence of deconfined matter, interactions of the colored jet charges with the medium constituents induce further out-of-cone energy loss contributions to the jet spectra. The size of these contributions depend on the interplay between the partonic structure contained within cone $R$ and the QGP properties. Accounting for these medium effects requires extending angular resummation to scales smaller than $R$, as described in the present Subsection.

The cross-section to measure a jet with momentum $p_T$ contained within a cone $R$ in AA collisions is expressed as
\begin{align}
\label{eq:sigma-AA}
\sigma^{\tiny AA}(\pT,R) &= \sum_{i=q,g} \int_0^\infty \rmd \epsilon \, P_i(\epsilon,R,p_T) \tilde \sigma^{\tiny pp}_i(\pT+\epsilon,R) \,, \nn
&= \sum_{i=q,g} Q_i(\pT,R) \tilde \sigma^{pp}_i(\pT) \,,
\end{align}
where $\tilde \sigma_{i}^{\tiny pp}$ is the quark/gluon contribution to the total cross-section at angular scale $R$, as computed in Eq.~\eqref{eq:sigma-pp} 
(the tilde serves as a reminder that the proton PDFs are replaced by the nuclear PDFs EPS09 computed at LO~\cite{Eskola:2009uj}, i.e.  $f_{i/p} \to f_{i/A}$ in Eq.~\eqref{eq:sigma-pp}), and $P_i(\epsilon)$ represents the probability distribution of losing energy $\epsilon$ out of the jet cone. We define the flavor dependent resummed quenching factor (QF) as~\cite{Baier:2001yt}
\beq
Q_i(\pT,R)\equiv\int_0^\infty \rmd \epsilon \, P_i(\epsilon,R,p_T)\frac{\tilde \sigma_i^{\tiny pp}(\pT+\epsilon)}{\tilde \sigma_i^{\tiny pp}(\pT)} \, .
\eeq
In the limit of large power index $n$, we can use following the asymptotic expansion, valid for $\epsilon\ll p_T$ and $n\epsilon \sim \pT$,
\begin{align}
\frac{\tilde \sigma^{\tiny pp}(\pT+\epsilon)}{\tilde \sigma^{\tiny pp}(\pT)} &= \frac{\pT^{n(\pT)}}{(\pT + \epsilon)^{n(\pT+\epsilon)}} \nn
&= \rme^{-\nu(\pT) \epsilon}\bigg[ 1+{\cal O}(\nu_2(\pT) \epsilon^2)\bigg] \,,
\end{align} 
where the flavor subscript has been omitted for clarity, and where $\nu(\pT) = \frac{\rmd \ln \tilde\sigma^{pp}}{\rmd \pT}  $ 
and $\nu_2(\pT) = \frac{\rmd^2 \ln \tilde\sigma^{pp}}{{\rmd \pT}^2} $. 
For our purposes, we will approximate $\nu(\pT) \approx n(\pT)/\pT$.
Then, to first non-vanishing order in this expansion~\footnote{For the impact of higher-order terms see~\cite{Takacs:2021bpv}.}, one can identify the quenching factor (QF) with the Laplace transform of the energy loss probability $P_i(\epsilon,R)$,
\begin{align}
Q(\pT,R) &\simeq \tilde P\left(\frac{n}{\pT},R\right) \,,\nn
&\equiv \int_0^\infty \rmd \epsilon \, \left.P(\epsilon,R) \rme^{-\nu \epsilon } \right|_{\nu=n/\pT} \,.
\end{align}
The factorization \eqref{eq:sigma-AA} reduces trivially to the jet production cross section in the absence of final-state interactions, Eq.~\eqref{eq:sigma-pp}, when $P(\epsilon)=\delta(\epsilon)$ (hence $Q= 1$) and by replacing the nuclear PDFs by proton PDFs.

The amount of quenching experienced by a quark/gluon jet of energy $E$ and angular extent $R$, $Q_i(E,R)$, depends on the amount of sources of energy loss it contains. These sources are understood as the jet substructure fluctuations induced by the vacuum-like DGLAP evolution at scales below $R$. The finite spacetime extent of the medium requires knowledge of the spacetime evolution of these substructure fluctuations as well. A key ingredient in this picture is the quantum-mechanical formation time of an emission process, which essentially corresponds to the time it takes for an emission to decohere from its emitter, e.g. when $|qg\rangle\to |q\rangle+|g\rangle$ for the case of gluon emission off a quark. This formation time can be estimated using the Heisenberg uncertainty principle to be $\tform \sim (E/Q)/Q$~\cite{Dokshitzer:1991wu}, where $Q$ is the virtuality of the process (not to be confused with the QF $Q_i$) and $E/Q$ is a boost factor to the co-moving frame of the particle. More explicitly, when considering the transverse momentum squared of a splitting event that yields two offsprings with longitudinal momenta $z E$ and $(1-z) E$, respectively,  which reads $k_\perp^2 = z(1-z)Q^2$, its formation time can be expressed as $\tform \sim z(1-z)E/k_\perp^2$.

A given fluctuation will only contribute to the QF if it is formed within the medium, characterized with an extent $L$, so $\tform<L$. Another requirement arises from the physics of color coherence within the medium. By computing stimulated emission off a color connected dipole traversing deconfined matter, it was found that independent emissions off each of the legs of the antenna could only happen after a time $\tdecoh$~\cite{Mehtar-Tani:2011hma,Mehtar-Tani:2011vlz,Casalderrey-Solana:2011ule,Mehtar-Tani:2012mfa}, the de-coherence time. This is the time it takes for the members of a dipole to lose their color correlation via color rotations induced by interactions with the medium constituents. The decoherence time  $\tdecoh$ defines the time when the size of the dipole $r_\perp(t) = \theta t$ is of order the inverse transverse momentum scale accumulated by multiple scattering during $t$, i.e., $Q_\perp(t) \equiv \sqrt{\hat q t}$. Or equivalently, when $r_\perp(t)Q_\perp(t) \sim 1 $. Solving for $t$ reads the decoherence time, $\tdecoh \sim (\hat q \theta^2)^{-1/3}$\,. A given fluctuation will only contribute if $\tdecoh<L$. Also, naturally, one requires $\tform<\tdecoh$ for the splitting to be considered vacuum-like. When $\tform>\tdecoh$ the splitting is medium induced unless it is produced outside the QGP. 

The non-linear angular evolution that couples the resummed QF of quarks and gluons is then given by~\cite{Mehtar-Tani:2017web}
\begin{align}
\label{eq:collimator-eq}
\frac{\del Q_i(p,\theta)}{\del \ln \theta} &= \int_0^1 \dd z \,\frac{\alpha_s(k_\perp)}{2 \pi} p_{ji}(z) \Theta_\text{res}(z,\theta) \nn
&\times \left[Q_j(zp,\theta) Q_k((1-z)p,\theta) - Q_i(p,\theta) \right] \,,
\end{align}
where the resolved phase-space constraint encapsulates the requirements stated above, i.e. $\Theta_\text{res}(p,R) = \Theta(\tform < \tdecoh < L)$, where $\tform = 2/(\omega \theta^2)$ and $\tdecoh = (\hat q \theta^2/12)^{1/3}$. The requirement $\tdecoh<L$ can be translated into an angular constraint, as $\theta>\theta_c$, where the critical angle $\theta_c=(\hat qL^3/12)^{-1/2}$ is the minimum angle a dipole can have such that it is resolved by the medium \footnote{The factor $12$ in the definitions of both $\theta_c$ and $\tdecoh$ result from the exact from of the decoherence parameter for a singlet antenna, $\ln\Delta_{\rm med}(t)=-\hat q \theta^2 t^3/12 = -(\theta/\theta_c)^2=-(t/\tdecoh)^3$ \cite{MehtarTani:2010ma,MehtarTani:2011tz}.}. In terms of angular scales, Eq.~\eqref{eq:collimator-eq} encodes the ``collimator resummation'' evolution depicted in Fig.~\ref{fig:sketch} at scales below $R$. This evolution has no effect for angles below $\theta_c$, labelled as ``coherence'' in the sketch. 

Even if a jet had a trivial in-medium evolution (for instance if $\theta_c > R$, it would still lose energy, but only from the source associated with the total color charge of the multi-partonic system. This corresponds to the initial conditions of Eq.~\eqref{eq:collimator-eq}, and are called the bare quenching factors, computed as $Q_i(p,0) = Q^{(0)}_{\text{rad},i} (\pT) Q^{(0)}_{\text{el},i}(\pT) $, for radiative and elastic energy loss mechanisms, respectively~\cite{Baier:2001yt,Salgado:2003gb}. 
The radiative factor is related to the Laplace transform of the quenching weight, and given by 
\begin{equation}
    \label{eq:rad-qf0}
    \tilde P(\nu) = \exp\left[- \int_0^\infty \rmd \omega\, \frac{\rmd I}{\rmd \omega} \left(1-\rme^{-\nu \omega} \right)\right] \,,
\end{equation}
where $\rmd I /\rmd \omega$ represents the spectrum of medium-induced emissions and $Q^{(0)}_{\rm rad}(\pT) \equiv \tilde P(\nu=n/\pT)$. We discuss both contributions in detail below, cf. Eqs.~\eqref{eq:radeloss} and \eqref{eq:eleloss} for final results.

Let us continue with a brief reminder of the framework used in the present and previous work~\cite{Mehtar-Tani:2021fud} to compute medium-induced radiation. It is based on the so-called improved opacity expansion (IOE) scheme, where the relatively rare hard momentum exchanges are computed as perturbations on top of a sea of frequent multiple soft scatterings. It amounts to expanding around the harmonic oscillator. This framework naturally incorporates the BDMPS-Z (multiple soft scattering) and GLV~\cite{Gyulassy:2000fs,Guo:2000nz} (single hard scattering) limits of the radiation spectrum, while it leaves out the Bethe-Heitler (BH) regime (corresponding to soft emissions with $\omega \lesssim T$)~\footnote{For recent developments in the analytical treatment of the BH regime together with the rest of regimes see~\cite{Isaksen:2022pkj}. We note that this regime, with radiated energies $\omega < T$, is of lesser phenomenological importance for the present work~\cite{Wiedemann:2000za,Andres:2020kfg}.}.

When the medium-induced emission is soft compared to the parent parton, i.e. $\omega \ll E$, the radiation spectrum can be expressed as~\cite{Mehtar-Tani:2019tvy,Mehtar-Tani:2019ygg,Barata:2020sav}  $\dd I_\text{NHO}/\dd \omega = \dd I^{(0)}/\dd \omega + \dd I^{(1)}/ \dd \omega$, where ``NHO'' (or next-to-harmonic oscillator approximation) refer to the order of the expansion in the IOE, and read
\begin{align}
\label{eq:dIdomega0}	
&\frac{\dd I^{(0)}}{\dd \omega} = \frac{2\alpha_s C_R}{\pi\omega} \, \ln \left|\cos \Omega L \right| \,, \\
\label{eq:dIdomega1}
&\frac{\dd I^{(1)}}{\dd \omega} = \frac{\alpha_s C_R \hat q_0}{2\pi}  \rmR \int_0^L \dd s\, \frac{-1}{k^2(s)}  \ln \frac{-k^2(s)}{Q^2\, \rme^{-\gamma_E}}  \,.
\end{align}
In the last expression, $\Omega = (1-i) \sqrt{\hat q /(4 \omega)}$, 
$k^2(s) =i \omega \Omega[ \cot \Omega s - \tan \Omega(L-s)]/2$, and the strong coupling constant runs with the typical transverse momentum of the emission, as $\alpha_s = \alpha_s\big((\hat q \omega)^{1/4} \big)$. These expressions are formally valid when the medium density is constant. We will nevertheless make use of them with re-scaled medium parameters to account for the dynamical expansion, see Sec.~\ref{sec:ebe} (for a discussion of the scaling properties, see \cite{Adhya:2019qse,Adhya:2021kws}).

In the IOE, the effective transport coefficient $\hat q$ acquires logarithmic corrections associated to the matching scale $Q$, differing from the bare $\hat q_0$ as
\beq
\label{eq:qhat-log}
\hat q = \hat q_0 \ln \frac{Q^2}{\mu_\ast^2} \,,
\eeq
where $Q$ naturally corresponds to the transverse momentum acquired by the emission via medium-interactions during its formation time. This scale is found by solving the implicit equation $Q^4 = \hat q_0 \omega \ln Q^2/\mu_\ast^2$, where the cutoff scale is $\mu_\ast^2 = m_D^2\, \exp[-2 + 2\gamma_E]/4 $ \cite{Mehtar-Tani:2019ygg,Barata:2020sav}. The bare jet transport coefficient follows is obtained within the Hard Thermal Loop (HTL) effective theory, and reads $\hat q_0 = \gmed^2 N_c m_D^2 T/(4\pi)$, while the expression for the Debye screening mass $m_D$, for three active flavours, reads $m_D^2 = 3\gmed^2T^2/2$. Finally, the coupling $\gmed$ between the jet partons and the medium constituents is taken to be fixed, and corresponds to the only free parameter of our framework.

While Eqs.~\eqref{eq:dIdomega0} and ~\eqref{eq:dIdomega1} capture the $\omega$-distribution of the emitted quanta correctly in the regimes of interest, they have been integrated over transverse momentum $k_\perp$ and offer no information about the broadening they experienced. In order to compute jet energy loss, one needs additional knowledge about their final angular distribution with respect to the jet axis, $\theta \sim k_\perp/\omega$, since only those emissions that end up outside of the jet cone $R$ contribute to the QF. To that end, we have adopted a multiplicative Ansatz that depends on the regime in frequency of the emission: hard or soft.

We distinguish gluons pertaining to three broad regimes. First, there are gluons with energies $T< \omega< \omega_s$, where $\omega_s \equiv (\gmed^2 N_c/(2\pi)^2)^2 \pi \hat q_0 L^2$, for which formally $\rmd I^{(0)}/\rmd \omega \gg \rmd I^{(1)}/\rmd \omega$ in the IOE expansion. These gluons are produced with $\mathcal{O}(1)$ probability during the passage of the parent parton through the medium and trigger an inverse turbulent cascade via democratic branchings that results into an efficient accumulation of modes around $ \omega \sim T$~\cite{Blaizot:2013hx,Blaizot:2014ula,Blaizot:2014rla,Iancu:2015uja}. Emissions pertaining to this regime are then assumed to have been thermalized, becoming part of the medium and contributing as correlated background. After the subtraction of the uncorrelated background, as it is typically done in experiments, their contribution to the final energy distribution will be that produced by the wake generated via the perturbation of the relativistic hydrodynamic equations of motion that govern the evolution of a liquid system such as the QGP. While realistically modelling this non-perturbative piece of dynamics is beyond the scope of the present work, we estimate its effects on the QF by considering the possibility that some of this energy can be recaptured. The larger the jet cone, the more likely it is to recapture some of the energy. 

Focusing first on the turbulent regime, corresponding to $T<\omega<\omega_s$, the amount of lost energy \emph{at any angle} can be estimated as
\beq
\Delta E_{\rm turb} = \int_T^{\omega_s}\rmd \omega \,\omega \frac{\rmd I^{(0)}}{\rmd \omega} \,.
\eeq
Now, assuming that a fraction $\xi$ of this energy gets thermalized and redistributed back inside the cone, just $(1-\xi) \Delta E_{\rm turb}$ is actually lost. Using geometrical arguments this ratio $\xi$ should correspond to the ratio of the area covered by the jet-cone and the area where the energy is distributed, arriving at $\xi \approx R^2/R^2_{\rm rec}$, where the parameter $R_{\rm rec}$ quantifies how efficient opening up the cone is in recapturing the thermalized energy. Assuming a flat angular distribution across the hemisphere of a given jet would correspond to $R_{\rm rec}=\pi/2$, while a narrower angular distribution, as inspired by linearized hydrodynamics wake calculations in the absence of radial flow, would correspond to $R_{\rm rec}< \pi/2$. Inspired by the study in \cite{Mehtar-Tani:2021fud}, we have currently chosen $R_{\rm rec}=(5/6)\pi/2$. In our previous publication we checked that the choice of this non-perturbative parameter has very little influence on jet suppression up to $R\lesssim 0.6$, even for extreme values of $R_{\rm rec}$, specially when compared to other sources of uncertainties, such as the determination of the resolved phase-space (which is perturbatively calculable). For this reason, and for jet observables up to $R\lesssim 0.6$, $\gmed$ still remains as the only relevant free parameter. 

In order to realize this smearing, we note that $\Delta E = - \frac{\rmd }{\rmd \nu}\tilde P(\nu) |_{\nu=0}$, see Eq.~\eqref{eq:rad-qf0}. Hence, we define a ``turbulent'' quenching factor $Q^{(0)}_{\rm rad, turb}(\pT) = \tilde P_{T<\omega<\omega_s}\left(\nu(1-R^2/R^2_{\rm rec} \right)$, where we have indicated that the integral over $\rmd \omega$ in \eqref{eq:rad-qf0} is constrained by the thermal scale and $\omega_s$. For reference, the full bare quenching factor is given in Eq.~\eqref{eq:radeloss}.
In summary, emissions between $T<\omega< \omega_s$ are assumed to thermalize quickly, and their angular distribution is approximately flat in the hemisphere of the jet.

Second, semi-hard gluons are those emitted with $\omega_s < \omega \lesssim \omega_c$, where the critical energy $\omega_c \equiv \hat q_0 \ln(\hat q_0 L / \mu_\ast^2) L^2/2$ is the maximal energy that a medium-induced emission produced by multiple soft scatterings can have. Third, harder gluons, with $\omega > \omega_c$, can be produced via rare single large momentum transfers and we have now $\rmd I^{(0)}/\rmd \omega \ll \rmd I^{(1)}/\rmd \omega$. Semi-hard gluons will typically experience re-scattering within the medium via further frequent transverse kicks, resulting in a diffusion process that features a Gaussian probability distribution. While harder gluons will also experience Gaussian broadening, they are already produced with a relatively large $k_\perp \gg \hat{q} L$ (since $\tform<L$), and a characteristic power-law tail $\sim \hat q_0 L/k_\perp^4$. The relative size of these contributions in the computation of the QF can be obtained by integrating the broadening distribution computed using the IOE framework~\cite{Barata:2020rdn} for $k_\perp$ scales larger than $w R$, i.e. $B\big(\omega R; Q_{\rm broad}^2 \big) = (\rmd I/\rmd \omega)^{-1} \int_{(\omega R)^2}^\infty \rmd k_\perp^2 \, \rmd I/(\rmd \omega \rmd k_\perp^2)$, where $Q_{\rm broad}^2$ represents the characteristic broadening scale of the different mechanisms.

In this way, the full radiation spectrum for emissions that end up beyond the jet cone, under this multiplicative Ansatz, reads~\cite{Blaizot:2014rla}
\begin{align}\label{eq:outofcone-spectrum-final}
\frac{\dd I_>}{\dd \omega} &=  B\big(\omega R; Q_s^2/2 \big) \frac{\dd I^{(0)}}{\dd \omega} \nn
&+  B\big(\omega R; \max \big[Q_s^2,16 \omega/(\pi^2 L) \big] \big)\frac{\dd I^{(1)}}{\dd \omega} \,.
\end{align}
In the LO term, the broadening scale was set to $Q_{\rm broad}^2=Q_s^2/2$, where the saturation scale is $Q_s^2=q_0 \ln(\hat q_0 L / \mu_\ast^2)L$. This is the typical momentum acquired via multiple soft scatterings averaged over all possible production points between $0$ and $L$~\cite{Blaizot:2014ula,Blaizot:2014rla}. The NLO term considers the broadening scale that results into a larger amount of broadening, either via the emission process itself, or by further multiple soft scatterings. The choice of scales in the NLO term correctly reproduce the GLV limit at high $w$ and $k_\perp$, up to logarithmic factors. For more details on the steps here employed to incorporate broadening we refer the reader to the Appendix A of our previous work~\cite{Mehtar-Tani:2021fud} and to~\cite{Blaizot:2012fh}.

The bare quenching weight associated with medium-induced radiation is then
\begin{align}
\label{eq:radeloss}
& Q^{(0)}_{\rm rad}(\pT) = \exp \Bigg[ -\int_{\omega_s}^\infty \dd \omega \, \frac{\dd I_>}{\dd \omega} \left(1 - \rme^{-\nu\omega} \right) \nn
& - \int_{T}^{\omega_s} \dd \omega \, \frac{\dd I^{(0)}}{\dd \omega} \left(1 - \rme^{-\nu\omega(1-\left(\frac{R}{R_\text{rec}} \right)^2)} \right) \Bigg] \,,
\end{align}
where we have approximated $\dd I_{\rm NHO}/\dd \omega \simeq \dd I^{(0)}/\dd \omega$ in the turbulent regime $T<\omega <\omega_s$ . The bare quenching factor for elastic energy loss is
\beq
\label{eq:eleloss}
Q^{(0)}_{\rm el}(\pT) = \exp\left[- \hat e L \nu \left(1-\left(\frac{R}{R_\text{rec}} \right)^2 \right) \right] \, .
\eeq
This expression is simply obtained as the Laplace transform of $\delta(\epsilon-\hat e L(1-R^2/R_\text{rec}^2))$. For a weakly-coupled plasma, as assumed in this work, the transport coefficient $\hat{e}$~\cite{Majumder:2008zg} is related to $\hat q$  by using the fluctuation-dissipation relation $\hat e = \hat q /(4T)$ (where $\hat e_q = \frac{C_F}{N_c} \hat e_g$ and $\hat e_g = \hat e$ for gluons) \cite{Moore:2004tg}.

\subsection{Observables}
\label{sec:observables}

The jet nuclear modification factor $R_{\rm \tiny AA}$ can be written as
 \begin{align}
     &R_{\rm \tiny AA}(\pT,R) = \frac{\sum_{i=q,g} Q_i(\pT,R) \tilde\sigma^{pp}_i(\pT,R)}{\sum_{i=q,g} \sigma^{pp}_i(\pT,R)} \\
     &= Q_q(\pT,R) \tilde f_q(\pT,R) + Q_g(\pT,R) \tilde f_g(\pT,R) \,.
 \end{align}
 As before, the tilded {\it pp} cross-section implies that the free-proton PDFs have been replaced by nuclear PDFs that account for nuclear modifications and isospin effects. In particular, the modified quark and gluon fractions, defined as
 \begin{equation}
     \tilde f_i(\pT,R) = \frac{\tilde \sigma^{pp}_i(\pT,R)}{\sum_{i=q,g} \sigma^{pp}(\pT,R)},
 \end{equation}
 do not add up to one, i.e. $\sum_{i=q,g}\tilde f_i(\pT,R) \neq 1$, reflecting some of the nuclear modifications that take place even in the absence of final state energy loss.

The other main observable studied in this work are the jet azimuthal anisotropies, and more specifically the so-called elliptic flow coefficient, $v_2$. It is defined as the second order Fourier coefficient in the expansion of the spectrum over the azimuthal angle $\phi$~\cite{Voloshin:1994mz,Poskanzer:1998yz}, and can be computed in AA collisions as
\begin{equation}
\label{eq:v2comp}
    v_2(\pT,R) \equiv \frac{\int_{-\pi}^{\pi}d\phi \cos(2 \phi) (d\sigma^{\rm AA}(\pT,R)/d\phi)}{\int_{-\pi}^{\pi}d\phi (d\sigma^{\rm AA}(\pT,R)/d\phi)} \, ,
\end{equation}
where we used that the uncertainty in the event-plane angle resolution $\Psi_R$ is negligible compared to other sources of uncertainties~\cite{ATLAS:2020yxw}, and we have set it to $\Psi_R=0$, as we do in our simulations.

\subsection{Event-by-event medium properties}
\label{sec:ebe}

In this Subsection we provide simple arguments that clarify the need to take into account fluctuations of the in-medium histories of jets in order to realistically compute jet observables in heavy-ion collisions. We also describe how we include them in our model.

Jets are produced at different locations in the transverse plane, with different orientations, and therefore explore different in-medium lengths and medium properties, such as temperature or flow, before escaping it. 
To clarify this point with an example, let us focus on the effect that fluctuations on the traversed length in the QGP, $L$, have on quenching.
One can define the average value of a given quantity $X$ by
\begin{equation}
    \langle X \rangle = \frac{\int_0^\infty \rmd L \, f(L) X (\pT,L)}{\int_0^\infty \rmd L \,f(L)} \, ,
\end{equation}
where $f(L)$ represents the probability to traverse a length $L$. In the most trivial event-by-event (EbE) scenario, we assume a flat distribution between values $L_{\rm min}$ and $L_{\rm max}$, yielding $f(L)_{\rm EbE}=\Theta(L-L_{\rm min})\Theta(L_{\rm max}-L)$. We compare the results obtained with this distribution against those obtained by computing the given quantity $X$ using only the average value (Ave) of $L$, where $f(L)_{\rm Ave}=\frac{L_{\rm max}+L_{\rm min}}{2}\delta(L-(L_{\rm max}+L_{\rm min})/2)$.

Let us now choose $X$ to be the amount of suppression of the spectrum. Again, for simplicity, let us just neglect the dependence on color factors and consider the bare quenching factor, ignoring thereby the phase-space resummation \footnote{In this simplified example, the resummed quenching weights give rise to expressions that are not particularly illuminating, so they will be omitted. Note that the deviations between the ``event-by-event'' and ``averaged'' scenarios are even larger than for the bare quenching weight case.}. Then, $X=Q=\rme^{-a L}$, where $a = \sqrt{2\pi \bar\alpha^2 \hat q \nu}$ with $\nu = n/\pT$. Here, we explicitly assume that $\hat q$ is constant throughout the medium and independent of time. The ratio of the mean quenching factor between the ``event-by-event'' and ``average'' scenarios is
\begin{equation}
    \frac{\langle Q \rangle_{\rm EbE}}{\langle Q \rangle_{\rm Ave}}=\frac{\rme^{-a L_{\rm min}}-\rme^{-a L_{\rm max}}}{a(L_{\rm max}-L_{\rm min}) \rme^{-\frac{a}{2}(L_{\rm max} + L_{\rm min}) }} \,.
\end{equation}
This ratio tends to 1 as $a$ goes to 0, for instance at high-$\pT$, but is larger than 1 at $a \gg 1$ \footnote{The ratio is enhanced at $\pT \ll n \bar\alpha^2 \hat q L^2$ where quenching effects are large , i.e., $Q\ll 1$.}. This difference can induce a change in the slope of $Q$ and can affect the agreement with high-precision experimental data in this regime. This very simple example illustrates that the average energy loss over many path-lengths is not in general the energy loss of the average path-length.

This discrepancy is not necessarily that striking for all quantities, though, as it obviously depends on the functional dependence of that quantity on $L$. For instance, in the case of $v_2$, the EbE result coincides exactly with the Ave. (in the small eccentricity approximation), but only if one uses the bare quenching weight,
\begin{align}
    \langle v_{2,i} \rangle_{\rm EbE} &= -\frac{2 e}{L_{\rm max}-L_{\rm min}} \int_{L_{\rm min}}^{L_{\max}} \frac{Q'_i}{Q_i} L dL \\ \nonumber
    &= \langle v_{2,i} \rangle_{\rm Ave} \, ,
\end{align}
simply because $Q'_i/Q_i$ is independent of $L$ \footnote{We used that $dL \equiv L 2 \epsilon$, where this $dL$ that quantifies the difference between $L_{\rm in}$ and $L_{\rm out}$ should not be confused with the $dL$ used to integrate over all possible $L$.}. This relation is no longer true when jet phase-space effects are included by making use of the resummed quenching factor because of its non-trivial $L$-dependence.

Therefore, in order to accommodate the variations in jet energy loss resulting from the diverse trajectories the jet may traverse within the expanding QGP, we must integrate our theoretical framework into a realistic heavy-ion environment. We follow the same procedure carried out in our previous publication~\cite{Mehtar-Tani:2021fud}, whose steps we summarize here for the reader's convenience.

We first determine the production point in the transverse plane $(x, y)$ by using the overlap of the thickness functions of the colliding nuclei, $T_{AB}(x, y; b) = T_A(x - b/2, y) T_B(x + b/2, y)$, with $b$ being the impact parameter of the nuclear collision. The thickness function is derived from the transverse density of nucleons within the Lorentz-contracted nuclei and is distributed following the Woods-Saxon density function~\cite{Miller:2007ri}. Then, we randomly assign an orientation within the transverse plane and a random rapidity value within the range $-2\leq y \leq 2$. While tracing the trajectory of the jet within the QGP, we calculate the integrated values of the necessary physical variables, which typically depend on the local temperature $T$ and fluid velocity $u$ until the jet exits the QGP phase, occurring at a pseudo-critical temperature that we choose to be $T_c=145$ MeV. These values of $T$ and $u$ are extracted from event-averaged hydrodynamic profiles~\cite{Shen:2014vra} that describe the evolution of an expanding droplet of liquid QGP in PbPb collisions at $\sqrt{s}=5.02$ ATeV, considering various centrality classes. Because parameters such as the fluid temperature are expressed in the local fluid rest frame, it is necessary to take into account the infinitesimal distance $\rmd x_F$ covered by the jet in this particular frame during each time increment $\rmd t$ in the laboratory frame.

Ignoring numerical constants, the essential physical variables required are obtained through integration along the trajectory $\gamma(t)$ of a jet as follows:
\begin{align}
L &= \int_{\gamma(t)} \rmd x_F \, , \nn
T &= \frac{1}{L} \int_{\gamma(t)} \rmd x_F\, T(x) \, , \nn
m_D^2 &\propto \frac{1}{L} \int_{\gamma(t)} \rmd x_F\, T^2(x) \, , \nn
\hat{q}_0 &\propto \frac{1}{L} \int_{\gamma(t)} \rmd x_F\, T^3(x) \, \left( \frac{p\cdot u(x)}{p^0} \right) \, , \nn
\hat{e} &\propto  \frac{1}{L} \int_{\gamma(t)} \rmd x_F\, T^2(x) \, \left( \frac{p\cdot u(x)}{p^0} \right) \, , \nn
\omega_c &\propto 2 \int_{\gamma(t)} \rmd x_F\, L(t) \, T^3(x) \, \left( \frac{p\cdot u(x)}{p^0} \right) \, , \nn
\theta_c^{-\frac{1}{2}} &\propto 3 \int_{\gamma(t)} \rmd x_F\, L^2(t) \, T^3(x) \, \left( \frac{p\cdot u(x)}{p^0} \right) \, ,
\end{align}
where the length $L(t)$ is the in-medium path of the jet up to time $t$. The dilution factor $(p\cdot u)/p^0$, with $p$ the four-momentum of the jet and $u$ the fluid four-velocity, arises in transport coefficients due to a flowing medium~\cite{Baier:2006pt}. We want to make clear that even though we account for fluctuations in the jet production point and orientation, we do not encode any additional fluctuations associated to the variations of the relevant quantities along the path of the jet.

\setlength{\tabcolsep}{11pt}
\renewcommand{\arraystretch}{1.5}
\begin{table}[t!]
    \centering
   \begin{tabular}{c|c|c}
    & \multicolumn{2}{c}{$\theta_c^\ast$} \\
    \hline
     Centrality & RHIC & LHC \\ \hline 
        0-5\% & 0.13 & 0.09 \\
        5-10\% & 0.15 & 0.10 \\
        10-20\% & 0.17 & 0.12 \\
        20-30\% & 0.22 & 0.15 \\
        30-40\% & 0.27 & 0.19 \\
        40-50\% & 0.35 & 0.24 \\
        50-60\% & 0.45 & 0.32 \\
        60-70\% & 0.58 & 0.41 \\
    \end{tabular}
    \caption{Fitted values for $\theta_c^\ast$, using Eq.~\eqref{eq:fitthetac}, both at RHIC (AuAu collisions at $\sqrt{s}=0.2$ ATeV) and LHC (PbPb collisions at $\sqrt{s}=5.02$ ATeV), for different centrality classes.}
    \label{tab:thetac}
\end{table}

\begin{figure}[t!]
    \centering
    \includegraphics[width=0.49\textwidth]{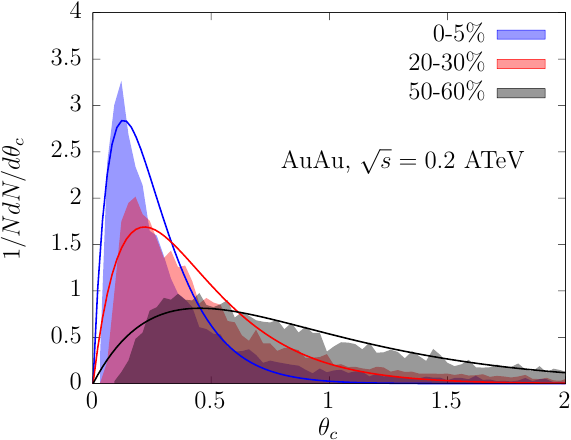}
    \caption{Probability distribution of $\theta_c$, as determined by the jet embedding procedure outlined in Section~\ref{sec:ebe}. We have just selected a few centrality classes for the collision kinematics at RHIC. The solid lines represent the fits obtained using Eq.~\eqref{eq:fitthetac}.}
    \label{fig:thetac_dists}
\end{figure}

Central to the physics discussion of this work is the length-dependence of the coherence angle $\theta_c$. In Fig.~\ref{fig:thetac_dists} we show the distribution of the values of $\theta_c$ at RHIC energies, for AuAu collisions, for a few centrality classes. To obtain these distributions we have followed the procedure outlined above. Given that the impact parameter increases towards more peripheral collisions, the traversed lengths tend to decrease notably. This means that the typical value of $\theta_c$ increases with decreasing centrality. This is precisely what is seen in Fig.~\ref{fig:thetac_dists}. Due to the reduced center-of-mass energies used at RHIC, the initial temperature of the QGP is lower, and therefore it reaches $T_c$ earlier, reducing the typical traversed length $L$. This implies that, for a given centrality class, the typical values of $\theta_c$ at RHIC will be larger than those at LHC. One can easily see that the $\theta_c$ distributions from Fig.~\ref{fig:thetac_dists} are of the type
\begin{equation}
\label{eq:fitthetac}
P(\theta_c)=\frac{\theta_c}{\theta_c^{\ast\,2}}\rme^{-\theta_c / \theta_c^\ast} \, ,
\end{equation}
where $\theta_c^\ast$ is the {\it mode}, i.e. the most frequent value, of the $\theta_c$ distribution for a given centrality class in a given collision system. This distribution has its origin in the distribution of path-lengths explored by the different in-medium jet histories, as well as in the distribution of values of $\hat{q}$, for each centrality class and collision system. The value of $\theta_c^\ast$, both at RHIC and LHC for a wide range of centrality classes, will be useful to  interpret the results, as discussed in Section~\ref{sec:discussion}. The qualitative agreement of this set of fits can be seen by comparing the solid lines with the data histograms in Fig.~\ref{fig:thetac_dists}, and the corresponding values of $\theta_c^\ast$ are tabulated in Tab.~\ref{tab:thetac}.
Further fluctuations associated to different initial state configurations 
will be studied in future work.

\begin{figure*}[t!]
    \includegraphics[width=1\textwidth]{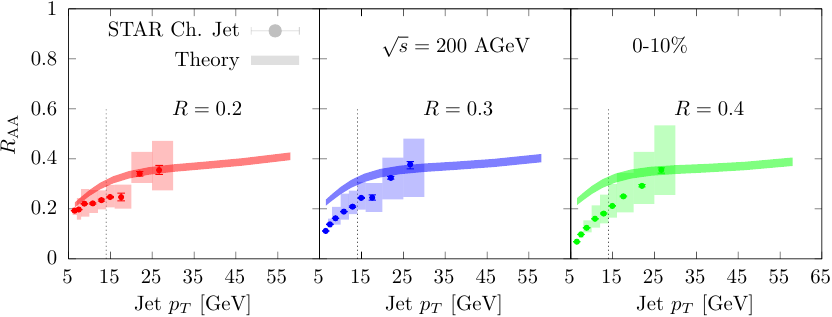}
    \caption{Comparison of jet $R_{\rm AA}$ against STAR data~\cite{STAR:2020xiv} measured at RHIC at $\sqrt{s}=0.2$ ATeV, where different panels show different $R$. The dashed vertical line indicates the jet $\pT$ below which the existence of an experimental bias prevents a direct comparison with our semi-analytical results.}
    \label{fig:dataraarhic}
\end{figure*}

\begin{figure}[t!]
    \includegraphics[width=0.5\textwidth]{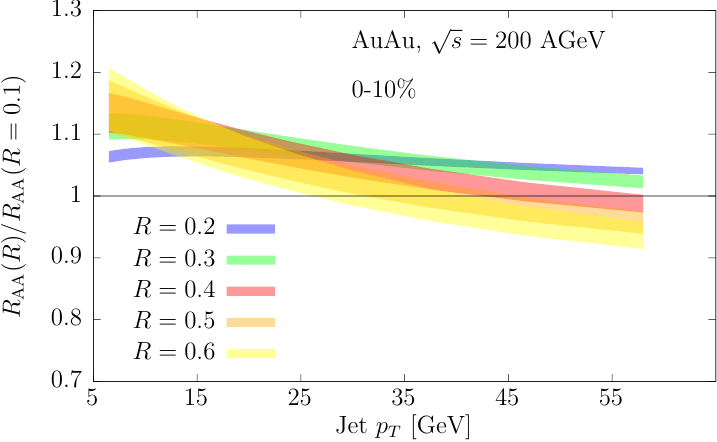}
    \caption{Ratio of $R_{\rm AA}$ for different $R$ over that for $R=0.1$, at $\sqrt{s}=0.2$ ATeV for AuAu collisions in the $0-10\%$ centrality class.}
    \label{fig:rhicraaratio}
\end{figure}

\section{Results}
\label{sec:results}

\begin{figure*}[t!]
    \includegraphics[width=1\textwidth]{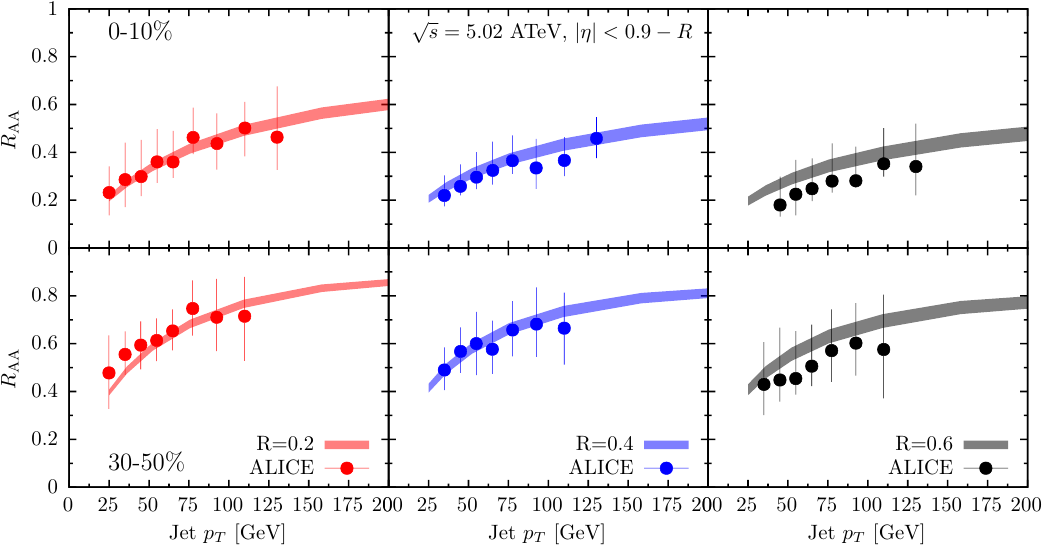}
    \caption{Comparison of theoretical predictions of $R_{\rm AA}$ against ALICE data~\cite{ALICE:2023waz} for PbPb collisions at $\sqrt{s}=5.02$ TeV, for the 0-10\% (30-50\%) centrality class in the upper (lower) row. Different columns display different values of jet-cone $R$.}
    \label{fig:dataraaalice}
\end{figure*}

In this Section we provide results for jet suppression, $R_{\rm AA}$, and jet azimuthal ansisotropy, $v_2$, as a function of the jet cone angle $R$ and jet transverse momentum $\pT$ for a number of centrality classes that range from 0-5\% to 60-70\%, both at RHIC and LHC. They are computed using the framework described in Section~\ref{sec:formalism}, which is the same as in our previous publication~\cite{Mehtar-Tani:2021fud}. We compare our predictions to experimental data that was in many cases released after the publication of our previous work. 

We emphasize that our calculation relies on two free parameters, $g_{\rm med}$ and $R_{\rm rec}$, namely the strength of the coupling to the medium, $g_{\rm med}$, and the energy recovery parameter, $R_{\rm rec}$ which were determined in our prior analysis of jet $R_{AA}$ at LHC energies \cite{Mehtar-Tani:2021fud}. Accordingly, in all of our results, the theory error bands correspond to taking the minimum and maximum values among the four combinations for the parameters $g_{\rm med}=\lbrace 2.2,2.3\rbrace$ and $R_{\rm rec}= \lbrace 1,5/6  \rbrace \frac{\pi}2$. 
We do not tune their values any further to obtain the results presented in the current work.

\subsection{Jet suppression at RHIC and LHC}
\label{sec:raa-predictions}
 
We start by showing in Fig.~\ref{fig:dataraarhic} out results for jet $R_{\rm AA}$ for three different radius, $R=0.2$, $R=0.3$ and $R=0.4$, in the different panels, confronted against experimental data measured by STAR~\cite{STAR:2020xiv}. Our results are computed for jets with $|\eta|<0.5$, while those from STAR have $|\eta|<1$.
Error bars on data denote statistical uncertainties, while bands refer to systematic uncertainties. The dashed line indicates the region below which the experimental results are biased by the requirement of the presence of a high-$\pT$ track within the jet, which is something that the STAR collaboration needed to apply in order to reduce the contamination from fake background jets, and that cannot be replicated by our present semi-analytical framework. Another difference is the fact that STAR only used charged tracks to reconstruct the jets, while our calculation is done for full jets. A rather crude way to account for this difference, which we did not attempt to do, would be to shift the jet $\pT$ by a factor $2/3$, although this would have little effect on the unbiased region due to the apparent relative flatness of $R_{\rm AA}$ with jet $\pT$. We observe good agreement between our predictions and experimental data in the unbiased region, above $\pT \gtrsim 15$ GeV, correctly reproducing the small dependence of jet suppression on jet-size $R$.

\begin{figure}[t!]
    \includegraphics[width=0.5\textwidth]{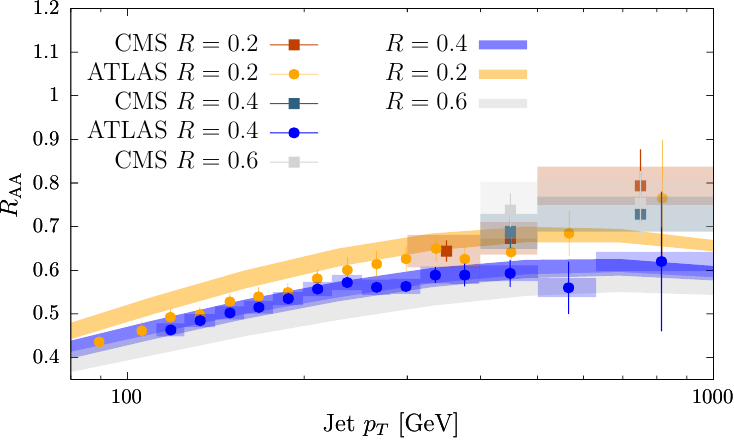}
    \caption{Our predictions for $R_{\rm AA}$ with $R=0.2$ confronted against ATLAS data~\cite{ATLAS:2023hso} and CMS data~\cite{CMS:2021vui}, for PbPb collisions at $\sqrt{s}=5.02$ TeV. In our previous publication~\cite{Mehtar-Tani:2021fud}, we used the first point of ATLAS data for $R=0.4$~\cite{ATLAS:2018gwx} for 0-10\% centrality class to fit our value of $g_{\rm med}$}
    \label{fig:dataraaatlas}
\end{figure}

In order to have a closer look into this mild $R$-dependence of jet suppression, we present ratios of different $R_{\rm AA}(R)$ over the result for $R=0.1$ for RHIC energy in Fig.~\ref{fig:rhicraaratio}. At low $\pT$, increasing the jet radius translate into less suppression (larger $R_{\rm AA}$), since the emissions off the very few resolved color charges can still be captured within the larger cone. As one increases the jet $\pT$, the phase-space within the jet increases, more so for a larger $R$ such as $R=0.6$, and the number of resolved charges increases. This leads to a larger suppression (still $\mathcal{O}(10\%)$ effects at most) of larger $R$ at higher $\pT$, overcoming the capturing effect that explains the behavior at low $\pT$. The $R$-dependence of jet suppression is thus sensitive to very relevant aspects of the jet-medium interaction, such as the distribution of the radiated energy and the size of the resolved phase-space. Future jet measurements at the sPHENIX detector~\cite{Belmont:2023fau} are expected to be precise enough to corroborate this picture.

After the publication of our previous work~\cite{Mehtar-Tani:2021fud}, the ALICE collaboration presented their results for the $R$-dependence of jet suppression at fairly low jet $\pT$, ranging from $R=0.2$ to $R=0.6$~\cite{ALICE:2023waz} (the $R-$dependence of this low jet $\pT$ range was first measured years ago by ATLAS~\cite{ATLAS:2012tjt}, although in terms of a double central-to-peripheral, $R_{\rm CP}$ ratio that we do not attempt to reproduce in the present work). In order to account for the differences in acceptance with respect to ATLAS' (whose results we compared against in our previous publication~\cite{Mehtar-Tani:2021fud}), i.e. $|y|<2.8$, we redid our calculations by modifying the rapidity ranges of the initial spectrum at $R_0=1$ from which we compute the vacuum and medium jet evolution, such that $|\eta|<0.9-R$. Any expected causes for differences in $R_{\rm AA}$ due to the different rapidity ranges, such as the change of the spectral index $n$ (larger at larger $\eta$) and the quark-initiated jet fraction (larger at larger $\eta$) turned out to be very small. We show our predictions, confronted against ALICE data, in Fig.~\ref{fig:dataraaalice}, obtaining an excellent agreement.

In Fig.~\ref{fig:dataraaatlas} we show jet $R_{\rm AA}$, also at LHC, but for a much higher jet $\pT$. Our results are compared to recent ATLAS data for $R=0.2$~\cite{ATLAS:2023hso}, ATLAS data for $R=0.4$~\cite{ATLAS:2018gwx} and CMS data~\cite{CMS:2021vui} for $R=0.2,0.4,0.6$. The first point, at $\pT \approx 100$ GeV of ATLAS results for $R=0.4$ jets~\cite{ATLAS:2018gwx}, was used to fit our value of $g_{\rm med}$ in \cite{Mehtar-Tani:2021fud}. The small larger suppression experienced by $R=0.4$ jets as compared to $R=0.2$ jets, which tends to disappear with decreasing jet $\pT$, as measured by ATLAS, is correctly captured by our prediction. As predicted in our previous work~\cite{Mehtar-Tani:2021fud}, jet $R_{\rm AA}(R=0.2)$ should be somewhat less suppressed than those with larger $R$, which hold even more similar values of suppression among them. To a large extent, this is so due to the fact that the typical critical angle for this centrality is around $\theta_c\sim 0.12$, representing an extensive part of the phase-space of an $R=0.2$ jet. However, this mild trend was not observed by CMS, where $R=0.2$ and $R=0.4$ lie on top of each other for the limited available range in jet $\pT$. Moreover, CMS has measured that also $R=0.6$ jets are as little suppressed as $R=0.2$ jets, together with $R=0.4$ jets, within current uncertainties. The manifest, although mild tension between the ATLAS and CMS results for $R=0.4$ jets motivates extending the $\pT$ range and precision of measurements. Further improvements from the theory side involve increasing the precision of the resolved phase-space calculation, encapsulated in the value of $\theta_c$ in our framework, as it can induce $\mathcal{O}(20\%)$ changes in jet suppression~\cite{Mehtar-Tani:2021fud}, specially at very high $\pT$.

In summary, the excellent agreement as a function of jet $\pT \gtrsim 100$ GeV up to $\pT \lesssim 1$ TeV, for $R=0.4$ jets belonging to a wide range of centrality classes shown in our previous work~\cite{Mehtar-Tani:2021fud}, combined with the no-less remarkable results here shown, which extend the $\pT$ reach down to $\pT\gtrsim 20$ GeV, both for smaller ($R=0.2$) and larger ($R=0.6$) jet radii, and for several centrality classes as well, present overall an encouraging picture of our first-principles understanding of jet suppression in deconfined QCD matter. Very good agreement is achieved across centralities, spanning almost two orders of magnitude in jet $\pT$, and several jet radii, up to $R=0.6$, which we deemed to be the radius above which non-perturbative effects due to correlated background (medium response) start to be important~\cite{Mehtar-Tani:2021fud}.

A complete set of predictions for jet $R_{\rm AA}$ at RHIC and LHC, for different centralities and jet radii up to $R=0.6$ can be found in Appendix~\ref{sec:fullscan}.

\subsection{Jet azimuthal anisotropy at RHIC and LHC}
\label{sec:v2-predictions}

We now turn to the discussion of the results for jet $v_2$, defined in Eq.~\ref{eq:v2comp}. In our main set of results we do not consider fluctuations in the initial collision geometry, as our hydrodynamic backgrounds are event-averaged within a given centrality class, and so we have set the event plane to $\Psi_R=0$. Just as it is the case for the flow coefficients of the soft particles from the bulk of the system, these fluctuations are expected to affect the values of higher-order flow harmonics, such as $v_3$, for high-$\pT$ objects as well~\cite{Noronha-Hostler:2016eow}. 


\begin{figure*}[t!]
    \includegraphics[width=1\textwidth]{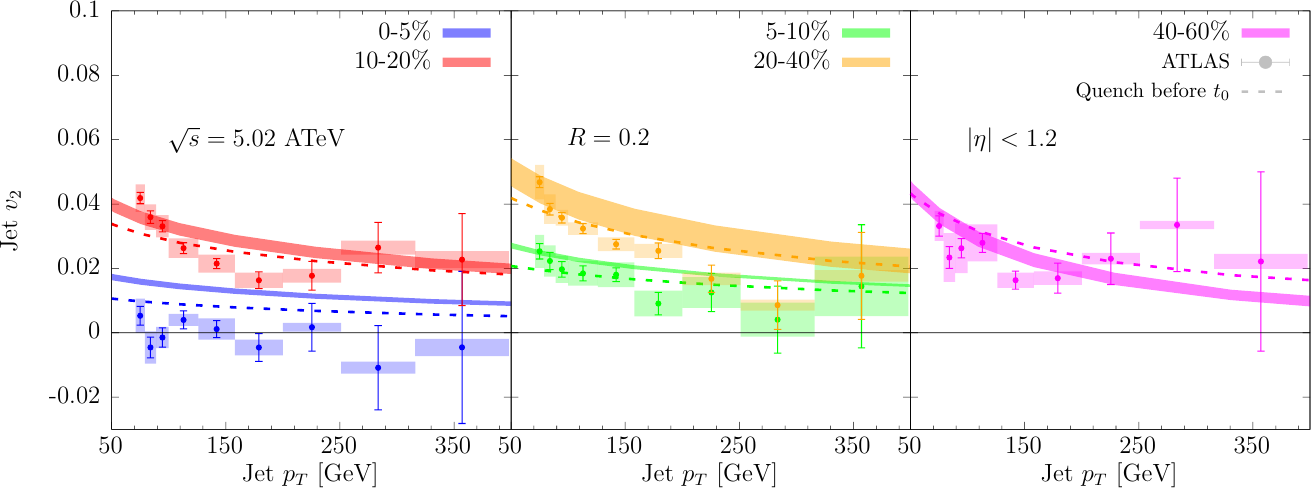}
    \caption{Results for jet $v_2(\pT)$ for $R=0.2$ jets at LHC for PbPb collisions at $\sqrt{s}=5.02$ ATeV, confronted against ATLAS data~\cite{ATLAS:2021ktw}. See text for details.}
    \label{fig:datav2lhc}
\end{figure*}

\begin{figure}[t!]
    \includegraphics[width=0.5\textwidth]{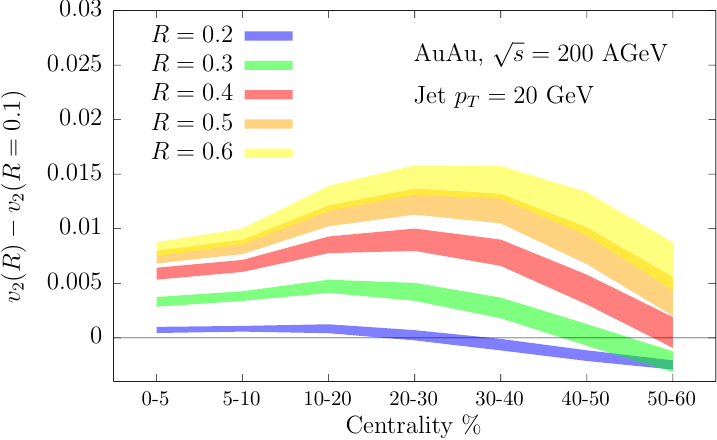}
    \caption{Difference between the jet $v_2(R)$ for $R>0.1$ with respect to $v_2(R=0.1)$, for jet $\pT=20$ GeV and as a function of centrality at RHIC for AuAu collisions at $\sqrt{s}=0.2$ ATeV.}
    \label{fig:rhicv2m0p1}
\end{figure}

In order to compute the jet $v_2$ as in Eq.~\eqref{eq:v2comp}, one simply needs to label each in-medium jet history with the azimuthal angle along which it propagated, $\phi$. One then readily obtains the results for $R=0.2$ jets at LHC, shown in Fig.~\ref{fig:datav2lhc}, which are compared against ATLAS data~\cite{ATLAS:2021ktw}. While ATLAS jet $R_{\rm AA}$ results typically feature an extended rapidity coverage, $|y|\leq 2.8$, these jet $v_2$ results are for jets within $|y|<1.2$. We have recomputed our double differential jet spectra, both in $p_T$ an $\phi$, for this narrower rapidity range. In contrast to the case of jet $R_{\rm AA}$, where we found negligible effects from narrowing the rapidity range when comparing to ALICE data, jet $v_2$ is more sensitive to the decrease of the quark-fraction that is induced by the reduction of the rapidity range about mid-rapidity, especially so towards lower jet $p_T$~\cite{Pablos:2022mrx}. The reason is that, in the soft limit, jet $v_2 \sim C_R$, as will be illustrated in Section~\ref{sec:discussion}. Our predictions (solid bands) yield in general very good agreement with experimental data, both in total magnitude and in the jet $\pT$-dependence. 

The comparisons shown in Fig.~\ref{fig:datav2lhc} point to the presence of tensions in the 0-5\% centrality bin, where we overpredict the size of jet $v_2$ at lower jet $p_T$. One possible reason could be our oversimplified treatment of quenching physics before hydrodynamization time, $t_0= 0.6$ fm/c in our simulations, where we simply ignore energy loss effects. It has been shown that neglecting quenching in the earliest stages overestimates the size of $v_2$~\cite{Andres:2019eus,Andres:2022bql}. Including the effects of quenching in the initial stages, for which theoretical calculations are now becoming available~\cite{Carrington:2022bnv,Boguslavski:2023alu}, will be studied in future work. In the present work, we estimate the effect by allowing the possibility of elastic and radiative energy loss before $t_0$, as if our present formulas were valid out-of-equilibrium as well. In practice, we set $T(t<t_0,\Vec{x})=T(t_0,\Vec{x})$ and $v_x(t<t_0)=v_y(t<t_0,)=0$, while still keeping $v_z=z/t$. Quenching during more time naturally reduces the value of jet $R_{\rm AA}$, so in order to test the effect on $v_2$ for the same amount of jet suppression, we need to reduce the value of the coupling with the medium. While the bands represent our default results with quenching $t\geq t_0$, obtained with $g_{\rm med}\in \lbrace 2.2,2.3\rbrace$, the dotted lines are the results with a reduced $g_{\rm med}=2$, so that jet $R_{\rm AA}$ for $R=0.4$ jets at 0-10\% centrality lies within the band. We observe that, with the exception of the most peripheral class, the dotted lines tend to be below the bands, yielding a smaller jet $v_2$, in consistency with what has been found in the literature~\cite{Andres:2019eus,Zigic:2019sth,Andres:2022bql}. We note that the largest relative effect is for the most central class, 0-5\%, where the reduction of jet $v_2$ is almost by a factor of two. While the reduction of jet $v_2$ does not significantly modify the level of agreement with experimental data for more peripheral classes, in the case of the 0-5\% class the sizeable relative reduction brings our theoretical predictions much closer to ATLAS data.

We further explore the $R$-dependence of jet $v_2$ in Appendix~\ref{sec:fullscan}, both at RHIC and LHC. In general, jet $v_2$ increases as a function of $R$, as expected, since a larger $R$ tends to imply a larger resolved phase-space. This is the main reason why $R_{\rm AA}$ tends to decrease with increasing $R \lesssim 0.6$ at high jet $\pT$. This trend tends to be reversed towards lower jet $\pT$, as illustrated for example in Fig.~\ref{fig:rhicraaratio}, where the effect of increasing the amount of recaptured energy by opening the cone overcomes the associated enlargement of the resolved phase-space. However, something distinct happens in the case of jet $v_2$, where always $v_2(R,\pT)>v_2(R',\pT)$ if $R>R'$, regardless of the relative ordering in $R_{\rm AA}$. This is best appreciated in Fig.~\ref{fig:raarhic}, for RHIC results, where the typically lower jet $\pT$ compared to LHC kinematics yield $R_{\rm AA}(R,\pT)>R_{\rm AA}(R',\pT)$, over a wide range in the accessible jet $\pT$, while $v_2(R,\pT)>v_2(R',\pT)$, for $R>R'$. 

In particular, we observe an interesting grouping among different $R$, evolving with centrality, which is again best perceived for RHIC kinematics. For central collisions, say 5-10\% centrality class, $v_2(R=0.1,\pT)\approx v_2(R=0.2,\pT)\lesssim v_2(R=0.3,\pT)<v_2(R=0.4,\pT)\lesssim v_2(R=0.6,\pT)$, while by the 50-60\% centrality class $v_2(R=0.1,\pT)\approx v_2(R=0.2,\pT)\approx v_2(R=0.3,\pT)\approx v_2(R=0.4,\pT)< v_2(R=0.6,\pT)$. This type of grouping is absent for $R_{\rm AA}$, and reflects the connection between jet $v_2$ and the path-length dependence of the critical angle $\theta_c$, whose most frequent value $\theta_c^*$ strongly evolves with centrality, as shown in Fig.~\ref{fig:thetac_dists}. These aspects are explained in Section~\ref{sec:discussion} based on simple arguments. At this stage, we merely illustrate the behavior of this grouping by computing the difference in jet $v_2$ between jets with a given $R>0.1$ and those with $R=0.1$, for a fixed jet $\pT=20$ GeV, as a function of centrality, as shown in Fig.~\ref{fig:rhicv2m0p1}. We observe the sequential collapse of the value of jet $v_2$ for jets with a given $R$ towards the value of jet $v_2$ for the smallest $R=0.1$ at different centrality classes. According to the projected uncertainties associated to jet $v_2$ measurements with the sPHENIX detector~\cite{sphenixBUP}, we believe that these subtle features can be confronted against upcoming RHIC data \footnote{Quite recently, preliminary measurements on jet $v_2$ have been shown by the STAR collaboration at RHIC~\cite{Sahoo:2023gho}, for ZrZr and RuRu collisions at 20-60\% centrality class. Due to the relative smallness of these systems, and the relatively peripheral centrality class selected, leading to fairly small typical values for path-length $L$, it is likely that these results are dominated by the $\theta_c > R$ region. However, despite uncertainties still being too large and the impossibility of our framework to reproduce the hard core matching selection performed in the experimental analysis, there are hints that $v_2(R=0.6)>v_2(R=0.4) \approx v_2(R=0.2)$ at around jet $\pT\approx 15$ GeV, which is qualitatively consistent with what is predicted in Fig.~\ref{fig:raarhic} for centrality classes more peripheral than 20\%.}.

\section{Discussion}
\label{sec:discussion}

To gain insight into the results presented in the previous Section, particularly to better comprehend the seemingly distinct behavior of jet $v_2$ concerning centrality compared to that of jet $R_{\rm \tiny AA}$, we will try to summarize the main features with an analytical model that captures the qualitative features of our fully-fledged event-by-event calculations above.

Our main goal is to reveal the importance of jet coherence, an aspect that has been largely neglected in existing literature. The geometry-sensitive nature of $v_2$ can be utilized to probe the length-dependence of both quenching \emph{and} resolution effects. To illustrate our point, consider measuring a quenched jet ``in-plane'' versus ``out-of-plane''. The jet traveling ``in-plane'' will not only experience a shorter path length, leading to a smaller bare quenching, but also be less resolved leading also to a larger $\theta_c$ and a smaller phase-space for additional substructure quenching. In contrast, the jet traveling ``out-of-plane'' is both more quenched \emph{and} more resolved. This additional effect emphasizes the path-length differences embodied in an observable such as $v_2$ and will lead to: \texttt{i)} an enhanced $v_2$ of jets compared to hadrons, and \texttt{ii)} a characteristic increase of $v_2$ with $R$ for a given centrality, as seen, for example, in Fig.~\ref{fig:rhicv2m0p1}. 

\subsection{$R$ dependence of $R_{\rm \tiny AA}$}
\label{sec:analytics-R-RAA}

Taking the strong quenching limit, i.e., $\Qbare_i\ll 1$, 
in Eq.~\eqref{eq:collimator-eq}, see \cite{Mehtar-Tani:2021fud} for details, and using the fact that the phase-space is dominated by the double log which allows us to approximate $z\sim 1$ (we further assume a fixed coupling constant), only the virtual term contributes and the solution can be written directly as
\begin{equation}
    \label{eq:qf-lin}
    Q_i(\pT,R) \simeq \Qbare_i(\pT)\,  \rme^{- \PS(\pT,R)} \,.
\end{equation}
Here, we effectively account the quenching of the leading parton (total charge) via $\Qbare_i$. For the moment, to simplify our discussion in this Section, we neglect any weak $R$-dependence of this factor. 
This phase-space is sensitive to resolved sub-jets in the QGP and is given by
\begin{align}
    \PS &= \int_0^R \frac{\rmd \theta}{\theta} \int_0^1 \rmd z \, \frac{\alpha_s}{\pi} P(z) \Theta(\tform < \tdecoh < L) \,,\nn
    &\approx  2 \bar \alpha \int_{\max(\theta_c,\theta_{\rm min})}^R \frac{\rmd \theta}{\theta} \int_{(\frac23 \hat q/\theta^4)^{1/3}}^{\pT} \frac{\rmd \omega}{\omega}\,,
\end{align}
where we used that the splitting function $P(z) \approx 2 C_R/z$ in the soft limit of gluon emission ($\omega = z \pT$). In this limit, we get that the formation time is $\tform = 2/(\omega \theta^2)$ and the decoherence time is $\tdecoh = (\hat q \theta^2/12)^{1/3}$. The limit on the angular integral is then $\theta_c = (\hat qL^3/12)^{-1/2}$ or $\theta_{\rm min} \equiv (\frac23 \hat q/\pT^3)^{1/4}$ \cite{Mehtar-Tani:2017web}. In the high-$\pT$ limt, precisely for $\pT > \omega_c /3$, and for $R > \theta_c$, we get
\begin{align}
    \label{eq:resolved-ps}
    \PS &= 2 \bar \alpha \ln \frac{R}{\theta_c} \left(\ln \frac{3 \pT}{\omega_c} + \frac{2}{3} \ln \frac{R}{\theta_c} \right) \,,
\end{align}
where the characteristic energy scale is $\omega_c = \hat q L^2/2$ and the characteristic angular scale is $\theta_c = (\hat q L^3/12)^{-1/2}$.
Clearly, if $R\leq \theta_c$ the phase-space vanishes $\PS = 0$ and the jet is quenched \emph{coherently}, only according to its total color charge, i.e.
\begin{equation}
    \left.Q_i(\pT,R)\right|_{R\leq \theta_c} = \Qbare(\pT) \,,
\end{equation}
where $\Qbare = \Qbare_\text{rad}\Qbare_\text{el}$, corresponding to radiative and elastic energy losses, respectively. We will not consider elastic energy loss in this Section, but note that most of the conclusions regarding the scaling properties of $R_{\rm \tiny AA}$ and $v_2$ still hold under the general condition that $\Qbare = \rme^{-f(\pT,R) L}$, where $f(\pT,R)$ is an arbitrary function of the jet transverse momentum and cone-angle.

For example, in what we refer to as the ``soft'' approximation, i.e. when assume that the medium-induced spectrum scales as $\rmd I /\rmd \omega \propto \omega^{-3/2}$, the bare quenching weight becomes $Q_i^{(0)}(\pT) = \exp( -2\sqrt{\pi \bar{\alpha}^2 \omega_c n/\pT})$ \cite{Baier:2001yt}, and only $\bar \alpha = \alpha_s C_i/\pi$ distinguishes between the quark and gluon energy loss.  This suppression factor describes how a single parton suffers energy loss through soft gluon emissions \emph{at any angle}. It becomes directly relatable to the quenching of single-inclusive hadron spectra if we replace the parton $\pT$ by the transverse momentum of the leading hadron, or $\pT \to \pT'\equiv \pT/\langle z \rangle$, see also \cite{Arleo:2017ntr}. For example, for massless hadrons $\langle z \rangle \approx 0.3-0.5$ over a wide jet $\pT$ range \cite{Sassot:2010bh}.

Assuming that the spectrum is dominated by a single parton species and neglecting nPDF effects, we find that
\begin{equation}
R_{\rm \tiny AA} \simeq Q(\pT,R)   \,.
\end{equation}
Including the full flavor dependence will influence jet $R_{\rm \tiny AA}$ (and hence also jet $v_2$) on the quantitative level only. Furthermore, for this discussion we omit effects coming from the recovery of emitted energy at large angles.

The regime of strong quenching, given by $\Qbare_i \ll 1$, is relevant for the mid- to high-$\pT$ range at RHIC and LHC. This is applicable in the mid- to high-$\pT$ regime where $\pT \approx 40$ GeV.
Hence, assuming the linearized form of the full quenching factor, cf. Eq.~\eqref {eq:qf-lin}, the ratio is
\begin{align}
    \frac{Q(\pT,R)}{Q(\pT,0.1)} & \approx 
    \exp\left[ \PS(\pT,0.1)-\PS(\pT,R) \right]\nn
    &\approx 1 - 2 \abar\mathcal{L}_c \ln\left( \frac{ \max (R, \theta_c) }{\max (0.1, \theta_c)} \right) \,,
    \label{eq:jet-raa-sketch}
\end{align}
where $\mathcal{L}_c\equiv \ln(3\pT/\omega_c)$,
up to terms that are not enhanced by a logarithm of $\pT$. We conclude that the ratio  of jet suppression factors is mainly logarithmically dependent on the coherence angle $\theta_c$.
 
\subsection{$R$ dependence of $v_2$}
\label{sec:analytics-R-v2}

Simplifying the precise details of the geometrical aspects of the suppression one can estimate the jet anisotropy as \cite{Zigic:2018smz}
\begin{equation}\label{eq:v2-def}
    v_2 \simeq \frac12 \frac{R_{\rm \tiny AA}(L_{\rm in})-R_{\rm \tiny AA}(L_{\rm out})}{R_{\rm \tiny AA}(L_{\rm in})+R_{\rm \tiny AA}(L_{\rm out})} \,,
\end{equation}
see also \cite{Adhya:2021kws}, 
where $L_{\rm in}$ is the shorter path-length experienced by the jets that are emitted in-plane, while $L_{\rm out}$ is the path-length of jets emitted perpendicular to it, i.e. out-of-plane. For slightly eccentric systems such as those corresponding to central to semi-central collisions we can assume that the typical differences in path-lengths between in-plane and out-of-plane are small, i.e. $L_{\rm in} = L$ and $L_{\rm out} = L + \Delta L$,
where $\Delta L/L \ll 1$. Assuming that the nuclear overlap resembles a perfect ellipse, the path length difference $\Delta L$ is directly related to the eccentricity of the ellipse $e$ via $\Delta L/L = 2 e$. It follows that \eqn{eq:v2-def} takes the simple form, 
\begin{equation}
    v_2 \approx -\frac{\Delta L}{4} \frac{{\rm d} \ln R_{_{AA}} }{{\rm d} L} \,.
\end{equation}
upon expansion to first order in $\Delta L$. 

To gain analytical insight into the behavior of $v_2$ as a function of jet $\pT$ and the cone angle $R$, we will at this point introduce further simplifications. Again, we assume that the spectrum is dominated by a single parton species and we will neglect nPDF effects. 
Furthermore, we will assume that the full quenching factor can be approximated by its linearized version with ``soft'' bare quenching factors, as done in the previous subsection, Sec.~\ref{sec:analytics-R-RAA}. These set of simplifications allow us to better isolate the effects of the relation between the resolved phase-space of a jet and energy loss.

In line with previous approximations, we expand the quenching weight up to the first order in $\Delta L$, i.e. $Q(L+\Delta L) \approx Q(L)(1+ \Delta L\, \delta Q)$ to obtain
\begin{equation}
    v_2 \approx -\frac{\Delta L}{4}\delta Q \,.
\end{equation}
We define the perturbed resummed quenching factor as $\delta Q(\pT,R) = \partial \ln Q(\pT,R)/\partial L$, where the path-length dependence is implicit through the dependence on the medium scales $\omega_c$ and $\theta_c$. The remaining part of this Section will be dedicated to studying the path-length dependence of the quenching factor through $\delta Q$ based on Eq.~\eqref{eq:qf-lin}.
As in the previous subsection, we will now focus on the regime of strong quenching, where $\Qbare(\pT) \ll 1$. This immediately implies $\delta Q \ll 1$. 

Let us start considering the path-length dependence of the bare quenching factor $\Qbare(\pT)$. We find that
\begin{equation}
\label{eq:deltaQ-bare}
    \delta Q_0= \frac{\partial \ln \Qbare(\pT)}{\partial L} \approx - C_R \sqrt{\frac{2 \alpha_s^2 \hat q n}{\pi\, \pT}} \,,
\end{equation}
where the last step is valid for the ``soft'' approximation.
Due to its specific dependence on the variable $\omega_c/\pT$, we can also trade the derivative on the path-length for a derivative on $\pT$, i.e. $\delta Q_0 = -\frac{2 \pT}{L} \frac{\partial \ln \Qbare(\pT)}{\partial \pT}$ \cite{Arleo:2022shs}.

Another contribution to the perturbation of the quenching factor resulting from the Sudakov suppression factor needs to be accounted for, i.e. $\delta Q = \delta Q_0 + \delta Q_1$, where $\delta Q_1 = \delta \PS(\pT,R)$, see Eq.~\eqref{eq:resolved-ps}.
This additional term encapsulates the path-length variation of additional jet splittings that occur in the medium. 
It turns out that the correction to the phase-space is
\begin{equation}
\label{eq:deltaPS}
    \delta \PS(\pT,R) = \frac{3 \bar \alpha}{L} \mathcal{L}_c \,,
\end{equation}
again with $\mathcal{L}_c \equiv \ln(3\pT/\omega_c)$,
which is independent of the jet cone angle up to the condition $R>\theta_c$.   
Although derived within simplified assumptions, the terms that we have identified allow us to extract several useful qualitative characteristics of jet $v_2$ that we can compare to the results from our full numerical simulations. Inspecting the two terms in Eqs.~\eqref{eq:deltaQ-bare} and \eqref{eq:deltaPS}, it becomes immediately clear that $v_{2,i}\propto \delta Q_i \propto C_R$ for $i=q,g$. This is simply a consequence of the Casimir scaling of the quenching factors, where both $\ln \Qbare_i \propto C_R$ and $\ln Q_i \propto C_R$ \footnote{For the resummed quenching factor, mild violations of Casimir scaling are implied by Eq.~\eqref{eq:collimator-eq}.}.

The flow coefficient is then found to be
\begin{equation}
    \frac{v_2}{e} 
    \simeq \bar{\alpha}\left(\sqrt{\frac{\pi \omega_c n}{\pT}}+\frac{3}{2}\mathcal{L}_c\,\Theta(R-\theta_c) \right) \,,
    \label{eq:jet-v2-sketc}
\end{equation}
where we highlighted the leading dependence of the second term on the cone-angle. In what follows, we will neglect the additional weak logarithmic dependence on $L$ in $\mathcal{L}_c$.
This illustrates a distinctive jump in jet $v_2$ at fixed $\pT$  and as a function of $R$ arising from the coherence physics encoded in the full quenching factor. Other sub-leading contributions to the $R$-dependence are not discussed further here  (see the comment in Sec.~\ref{sec:analytics-R-RAA}).

Note also that in the ``coherent regime'', i.e. for $R \leq \theta_c$, we uncover a simple relation between jet quenching and the azimuthal asymmetry,
\begin{equation}
    \left.\frac{v_2}{e}\right|_{R \leq \theta_c} \approx -\frac12 \left.\ln R_{\rm \tiny AA}\right|_{R\leq \theta_c} \,,
\end{equation}
expected to hold for modest path-length differences equivalent to small eccentricity. Assuming a relatively central collision, 20-30\%, for which $e\sim 0.2$ \cite{Eyyubova:2021ngi}, with a charged hadron $R_{AA}\sim 0.6$ around $p_T\approx 40$ GeV~\cite{CMS:2016xef} the above pocket formula predicts the estimated value $v_2\sim 0.05$, which is in the ballpark of the experimental data~\cite{CMS:2017xgk}. 
Note that this correspondence is exact in our model study, where we neglected expansion effects and employed the ``soft'' approximation. It should nevertheless hold under general assumptions as long as the ``bare'' quenching factor scales with $L$, i.e. $\ln \Qbare \propto L$. 
Finally, it is worth pointing out that the azimuthal asymmetry is a small $v_2/e \sim \mathcal{O}(\alpha_s)$ effect, in contrast to the overall jet suppression factor $\raa \sim \mathcal{O}(1)$.

The sensitivity to the critical angle $\theta_c$ is a direct consequence of coherence physics and therefore, a concrete prediction of our calculation. For example, instead of the coherent phase-space, if we assume that all emissions created in the medium, i.e. with $\tform < L$, fully contribute to jet quenching, we arrive at a collimator function with the phase-space $\PSincoh(\pT,R) = \frac{\bar \alpha}{2}\ln^2(\frac12 \pT R^2 L)$ \cite{Mehtar-Tani:2017web}, again for large enough $\pT$ so that the argument of the log is positive. In this case, there is a similar logarithmic enhancement of jet $v_2$ since $\delta \PSincoh = \frac{\bar \alpha}{L} \ln(\frac12 \pT R^2 L)$. However, this contribution is expected to appear for all $R$, leading to a much weaker relative $R$-dependence. 

Our above discussion also serves to draw a direct link between single-inclusive hadron and fully reconstructed jet $v_2$ at high-$\pT$. 
The hadron $v_2$ originates from a parton $v_2$ at a $\pT$ which is larger by a factor $1/\langle z \rangle $, and, since the quenching of the total charge contribution diminishes with $\pT$, it will always be smaller. On the other hand, jet $v_2$ is enhanced compared to the hadron $v_2$ due to the same reasons that $\raa$ is lower.

\begin{figure*}[t!]
    \centering
    \includegraphics[width=0.49\textwidth]{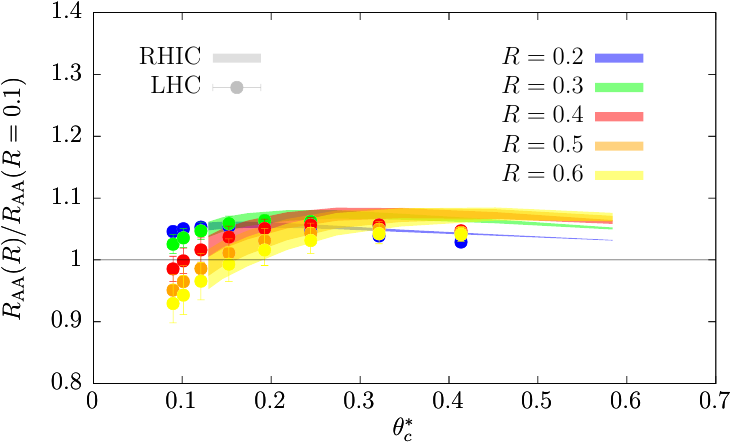}
    \includegraphics[width=0.49\textwidth]{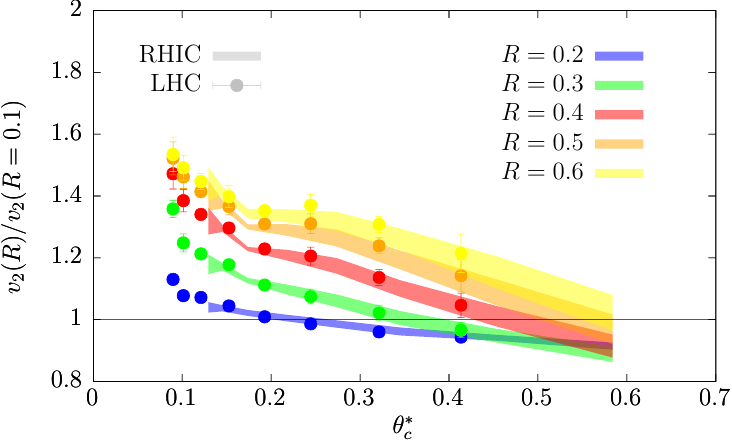}
    \caption{Left: Double ratio of $R_{\rm AA}$ between jets with different radii $R$ and jets with $R=0.1$. Right: Double ratio of jet $v_2$ between jets with different radii $R$ and jets with $R=0.1$. Results are plotted against the variable $\theta_c^*$, which is the most frequent value of the $\theta_c$ distribution for a given collision system and centrality. Jet momentum has been set to be $\pT \approx 38$ GeV.}
    \label{fig:prescaling}
\end{figure*}

\subsection{Identifying an angluar scale from jet suppression}
\label{sec:scaling-analytically}

The different qualitative behavior depending on the relative size of $R$ and $\theta_c$ found in previous sections suggests a representation of observables based on the ratio $\theta_c/R$. This approximate scaling behavior highlights the interplay between quenching and resolution effects affecting jets in the quark-gluon plasma.

In Section~\ref{sec:ebe} we have not only learned that the typical value of $\theta_c$ evolves strongly with centrality, but also that within a given centrality the corresponding $\theta_c$-distribution is relatively wide. We have extracted the most frequent value of $\theta_c$ for each centrality and collision system, $\theta_c^\ast(\sqrt{s},\rm{cent.})$, shown in Tab.~\ref{tab:thetac}. We will now make use of these set of values to present results in terms of the ratio $\theta_c^\ast/R$. 

We plot the jet $\raa$ and jet $v_2$ results obtained with the full semi-analytical calculations employed in Section~\ref{sec:results} in Fig.~\ref{fig:prescaling}. In order to factor out absolute effects related to the total amount of suppression, we have presented the results for jets with $R\geq0.2$ divided by the results for jets with $R=0.1$. Small-$R$ jets, such as $R=0.1$, are meant to be proxies for relatively unresolved color structures. The $x$-axis of these plots is a convenient measure of centrality, with 0-5\% being the most central and 60-70\% the most peripheral (as illustrated in Tab.~\ref{tab:thetac}). Given that $\theta_c$ decreases with path-length, we have chosen $\theta_c^\ast$ as such a measure (a smaller $\theta_c$ corresponds in general to more central collisions). In this plot, the data points represent calculations for LHC kinematics with model parameters identical to the analysis of the experimental data above. The shaded lines are equivalent calculations for RHIC kinematics. 

While the $\raa$ ratios for different cone-sizes $R$ are fairly similar to each other, with deviations $\mathcal{O}$(10\%), we see a strong ordering of the jet $v_2$ ratios with $R$, as expected from the qualitative discussion above. The jet $\pT$ has been set to be $\pT\approx 38$ GeV, since this is a regime accessible in both collision systems. Next, we note that the calculations for RHIC and LHC kinematics fall on top of each other. Focusing on the $v_2$ plot in Fig.~\ref{fig:prescaling} (right), we can see that the 20-30\% centrality class for LHC kinematics (corresponding to the third data point in the figure) overlaps with the prediction for RHIC kinematics in the 0-5\% centrality class (the leftmost part of the band). This reveals how the underlying distributions of $\theta_c$ are similar (see Table~\ref{tab:thetac} for the most frequent value $\theta_c^\ast$) and that this is driving the behavior of $v_2$.

\begin{figure*}[t!]
    \centering
    \includegraphics[width=0.49\textwidth]{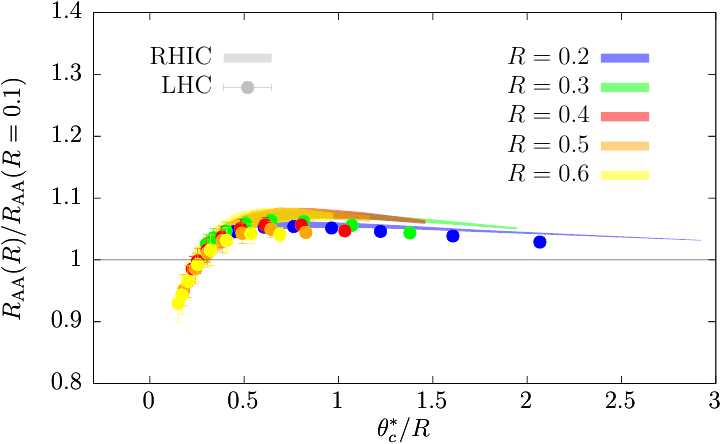}
    \includegraphics[width=0.49\textwidth]{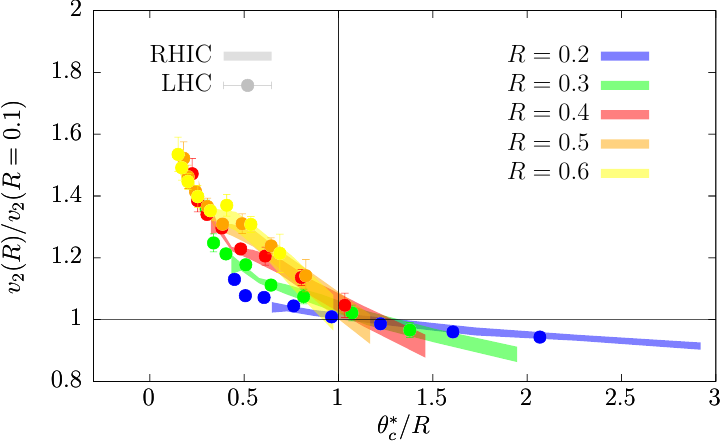}
    \caption{Left: Double ratio of $R_{\rm AA}$ between jets with different radii $R$ and jets with $R=0.1$. Right: Double ratio of jet $v_2$ between jets with different radii $R$ and jets with $R=0.1$. Results are plotted against the variable $\theta_c^*/R$, where $\theta_c^*$ is the most frequent value of the $\theta_c$ distribution for a given collision system and centrality. Jet momentum has been set to be $\pT \approx 38$ GeV.}
    \label{fig:scalings}
\end{figure*}

In Fig.~\ref{fig:scalings} we present results for jet $R_{\rm AA}$ (left panel) and jet $v_2$ (right panel) as a function of the scaling variable $\theta_c^\ast/R$. We can observe a striking collapse of all data points onto clearly defined paths along $\theta_c^\ast/R$, both for the jet $\raa$ and jet $v_2$ ratios shown in each panel. 
As above, the jet $\pT$ has been set to be $\pT\approx 38$ GeV, since this is a regime accessible in both collision systems.
Computations for RHIC kinematics reach larger values of $\theta_c^\ast/R$ than LHC kinematics, and LHC results reach smaller values of $\theta_c^\ast/R$ than RHIC results. 

Starting with the right panel of Fig.~\ref{fig:scalings}, where we plot the scaling of jet $v_2$ ratios, one of the most visible features is the congruence at $\theta_c^\ast/R\approx 1$. This is a clear manifestation of the collapse already hinted at in Fig.~\ref{fig:rhicv2m0p1}, and is expected based on the estimates from Eq.~\eqref{eq:jet-v2-sketc} for unresolved jets. Most importantly, this reveals a radical change of behavior of jet quenching at $\theta^\ast_c/R <1$, where jet substructure fluctations are contributing to the overall quenching, and $\theta^\ast_c/R >1 $, where the jets are quenched coherently. Succinctly, this plot illustrates the main features of our predictions for length-differential jet quenching at different center-of-mass energies, centralities and cone-sizes $R$.

The observed general trends can be easily understood using the simple analytical results of the previous Sections. As already pointed out in Sec.~\ref{sec:ebe}, attempting to compare this analytical toy-model to our complete framework requires at least the inclusion of some measure of the fluctuations of the dominant variables. Therefore, for a \emph{qualitative} comparison, we allow the fluctuation of $\theta_c$ while keeping all other parameters fixed (e.g. $\omega_c$). In this way, the step-like behavior of Eq.~\eqref{eq:jet-v2-sketc} now becomes similar to an exponential decay, i.e.
\begin{align}
\Theta(R-\theta_c) &\to \int_0^{R}\rmd\theta_c\, P(\theta_c) =1-\frac{1+t}{t\,\rme^{\frac{1}{t}}} \,,
\end{align}
where we used the parameterization in Eq.~\eqref{eq:fitthetac} and $t=\theta^\ast_c/R$. A similar procedure is performed to ``improve'' the toy-model of the ratio of jet suppression factors.

\begin{figure*}[ht!]
    \centering
    \includegraphics[width=0.49\textwidth]{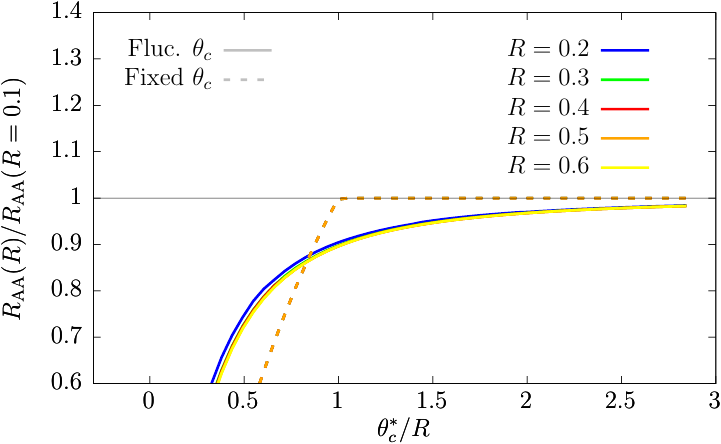}
\includegraphics[width=0.49\textwidth]{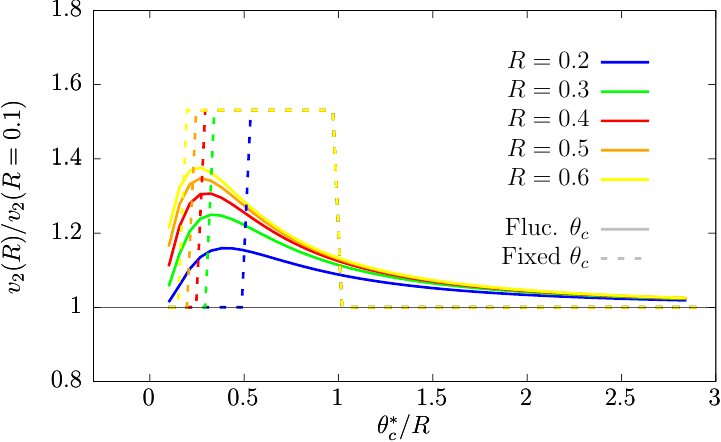}
    \caption{Simple analytical estimates for the jet $R_{\rm AA}$ (left panel) and jet $v_2$ (right panel) ratios between jets with different $R>0.1$ and those with $R=0.1$, at a fixed $\pT/w_c\approx 1.2$ and $n=5$. Results are presented as a function of $\theta_c^*/R$. The dashed curves labeled with ``Fixed $\theta_c$'' ignore the effect of $\theta_c$ fluctuations within a given centrality class, while the solid curves, labeled as ``Fluc. $\theta_c$'', do include them.}
    \label{fig:simplescalings}
\end{figure*}

Replacing the step function with this expression gives a very good approximation of the trend observed in Fig.~\ref{fig:scalings}. We plot the results using Eqs.~\eqref{eq:qf-lin} and \eqref{eq:jet-v2-sketc}, with the procedure to account for fluctuations in $\theta_c$ described above, in Fig.~\ref{fig:simplescalings}. For this qualitative comparison we have chosen $\pT/\omega_c\approx 1.2$ (roughly corresponding to $\widehat{q}=1.5 \, \rm{GeV}^2/\rm{fm}$, $L=3$ fm and $\pT=40$ GeV) and $n=5$. The dotted lines show the result had we kept $\theta_c$ at a constant value, while the full lines include fluctuations of $\theta_c$ according to Eq.~\eqref{eq:fitthetac}. As expected, fluctuations cause the scaling behavior of $v_2$ to drastically change shape, but the analytical trends are strikingly similar to the behavior obtained from the full numerical calculation.

Some remarks are in order. Note that in the left panel of Fig.~\ref{fig:simplescalings} the double ratio never goes above one in contrast to the left panel of Fig.~\ref{fig:scalings}, showing the full numerical results. Similarly, the toy-model predictions for the $v_2$ ratios in the right panel of Fig.~\ref{fig:simplescalings} do not go below one at $\theta_c^\ast/R >1$, in contrast to the right panel of Fig.~\ref{fig:scalings}. This is because our simplified formulas do not account for the possibility that radiation can stay within the cone, implying that larger $R$ always translates into a larger energy loss. In this coherent regime, the physics of jet quenching is dominated by the energy loss of a single color charge and we know that opening up the cone-angle leads straightforwardly to the slow recapturing of lost energy at large angles.

The fact that the clear jet $v_2$ scalings presented in the right panel of Fig.~\ref{fig:scalings} can be obtained simply by using the most frequent value of $\theta_c$ in a given centrality class, c.f. Tab.~\ref{tab:thetac}, is highly non-trivial. In comparison, the scaling observed for jet $R_{\rm AA}$ are not as striking given that jet suppression factor has a mild $R$-dependence to begin with, for a given centrality.

Despite all the differences associated to the quenching of jets with different $R$, produced in different collision systems, with mediums of different initial energy densities and lifetimes, to mention a few of them, we predict that these jet $\raa$ and jet $v_2$ ratios should follow general trends if expressed in terms of $\theta_c^*/R$. We reemphasize the fact that these trends are obtained from the results generated with the full semi-analytical framework, within a realistic heavy-ion environment, which successfully describe jet $\raa$ and jet $v_2$ for essentially all jet $p_T$ and $R$, centrality classes, and collisions systems -- after fitting a single parameter, $g_{\rm med}\approx2.25$.

\section{Conclusions}
\label{sec:conc}

In this paper we have provided the first analytical calculation of the elliptic flow coefficient $v_2$ for jets in heavy-ion collisions. Given that jets are extended multi-partonic objects, their degree of suppression due to medium interactions is in general larger than for single-inclusive hadrons. This fact leads, on average, to lower values of $R_{\rm AA}$, the jet yields suppression observable, and to larger values of $v_2$, the azimuthally asymmetric jet yields observable. A key ingredient of our framework is the determination of the quenched phase-space of a jet, determined by the physics of color decoherence in the medium. At LL accuracy, dipoles with an angle smaller than the critical angle $\theta_c$ will not be resolved by the medium, and will lose energy as a single color charge, i.e. will lose less energy. 
The fact that the critical angle possesses a marked length-dependence, $\theta_c \sim L^{-3/2}$, together with the fact that jets belonging to different centrality classes typically traverse very different amounts of QGP, naturally motivates the study of the physics of color decoherence for jet quenching observables as a function of centrality, targeting wide ranges of critical angles. The identification of such a program is one of the main conclusions of the present paper. 

As a relevant phenomenological example, we have studied the jet $v_2$ evolution with centrality as a function of the jet opening angle $R$. We have compared to ATLAS data for narrow jets with $R=0.2$, obtaining very good agreement, using the same framework and same single free parameter used in our previous work where we described the centrality evolution of inclusive jet suppression at LHC, for different jet radii~\cite{Mehtar-Tani:2021fud}. In contrast to the $R$-dependence of jet $R_{\rm AA}$, we have found that there is a clear ordering for the jet $v_2$ results for jets with different $R$, where larger $R$ jets always possess larger values of jet $v_2$ as compared to smaller $R$ jets. The reason is tightly connected to the typical value of $\theta_c$ in a given centrality class. Given that $v_2$ quantifies the different degree of suppression of jets traversing different amounts of length (i.e. it is a length-differential observable), those jets whose paths lead to a larger value of $\theta_c$ (the short paths) will be less suppressed than those whose paths lead to smaller values of $\theta_c$ (the longer paths), thereby raising the value of jet $v_2$. This is so unless the jet as a whole is to be unresolved regardless of the path taken, i.e. when $R<\theta_c$. In this case, small differences in the traversed path length will not lead to different amounts of resolved phase-space, reducing the size of jet $v_2$ when compared to those jets with $R>\theta_c$. We indeed observe how, as a function of centrality, jets whose $R$ goes from being above to below the typical value of $\theta_c$ in that centrality change their behavior in relation to jets with other $R$, such as in relation to the smallest $R$ jets studied in this work, $R=0.1$, as shown in Fig.~\ref{fig:rhicv2m0p1}.

Usage of a semi-analytical framework has allowed us to qualitatively explain, with simple expressions, this characteristic behavior of the $R$-dependence of $v_2$. These results suggest a representation of jet $v_2$ ratios in terms of the ratio between the two most relevant angular scales in this problem, i.e. as a function of $\theta_c/R$. In Fig.~\ref{fig:scalings} we have shown that this procedure leads to the approximate collapse of all such jet $v_2$ ratios, for any centrality and collision system, onto a single curve, for a fixed jet $\pT$. Our simple expressions, Eqs.~\eqref{eq:jet-raa-sketch} and \eqref{eq:jet-v2-sketc}, provide an analytical explanation for this behavior, provided that the substantial role of the $\theta_c$ fluctuations within a given centrality class are taken into account. The success of our framework to describe all presently available jet suppression data for any jet $\pT$, size $R$, centrality class and collision system grants robustness to these predictions. They represent a transparent test of the physics of color decoherence which can be confronted with future precise experimental data to be measured both at RHIC and LHC. This dynamical picture can be tested by acquiring precise data on jet $v_2$ for as many cone-sizes $R$ as possible, in as many centrality classes as possible and in as many collision systems as possible (such as PbPb, AuAu and OO collisions~\cite{Citron:2018lsq,Brewer:2021kiv}).

In this work we have also revisited the predictions for jet $R_{\rm AA}$, comparing our predictions with newly released experimental data and obtaining remarkably good agreement. Further directions include incorporating a running coupling between the jet parton and medium constituents, which is now fixed and is in practice the only free parameter of our framework ($g_{\rm med}$). Another particularly relevant theoretical improvement will consist in the inclusion of theoretically well-motivated dynamics for quenching in the non-equilibrium regime. Quenching in the initial stages can have sizeable effects on jet $v_2$, as shown in Fig.~\ref{fig:datav2lhc} using a simplified prescription, and in consistency with what was found for single-inclusive high-$\pT$ hadron $v_2$~\cite{Andres:2019eus,Andres:2022bql}. The strongest relative effect appears for the most central class, and is necessary to describe experimental data. 
Properly accounting for quenching in the initial stages is also necessary to understand the possible existence of energy loss in small systems, such as pPb or pp collisions.

The current implementation of medium response effects, now simply accounted for by a single parameter, $R_{\rm rec}$, and which have been shown to have important effects for large-$R$ jets, such as $R>0.6$, leaves room for improvement. 
A more precise treatment of the kinematics of the particles involved in the scattering processes, such as the recoiling constituents of the QGP, can be achieved by numerically computing bare quenching factors using QCD effective kinetic theory simulations~\cite{Mehtar-Tani:2018zba,Schlichting:2020lef}. 
Moreover, including the effects of thermodynamic gradients and flow on the stimulated radiation and broadening kernels recently computed in~\cite{Sadofyev:2021ohn,Barata:2022krd,Andres:2022ndd,Barata:2023qds,Kuzmin:2023hko}, so far neglected in most studies in the literature, will provide a more realistic description of the coupling between the jet and the dynamically evolving medium.
In particular, the existence of a flowing medium implies that the thermalized energy and momentum deposited by the jet will flow along with this medium, leading to different wake patterns depending on the jet in-medium history~\cite{Chaudhuri:2005vc,Tachibana:2014lja}. Large-$R$ jets are also sensitive to the amount of quenching experienced by the recoiling jet in the same event due to the long-range correlations induced by the wake~\cite{Yan:2017rku}, provided that they are close enough in rapidity~\cite{Pablos:2019ngg}. One way to account for these physics will be to use computationally efficient semi-analytical calculations of the wake profiles~\cite{Casalderrey-Solana:2020rsj,wakeprep} in order to determine the $R_{\rm rec}$ parameter jet by jet.

Furthermore, it will also be very interesting to extend this framework in order to go beyond inclusive jet observables, allowing it to becoming differential in jet substructure properties, or including the possibility of multi-jet events. 

\begin{acknowledgments} 
Computations were made on the B\'eluga supercomputer at McGill University, managed by Calcul Qu\'ebec and Compute Canada. 
Y. M.-T. was supported by the U.S. Department of Energy, Office of Science, Office of Nuclear Physics, under contract No. DE- SC0012704. D.P. has received funding from the European Union’s Horizon 2020 research and innovation program under the Marie Sklodowska-Curie grant agreement No. 754496, and is supported by European Research Council project ERC-2018-ADG-835105 YoctoLHC and by Spanish Research State Agency under project PID2020-119632GB- I00. 
\end{acknowledgments}


\input{main.bbl}


\appendix

\begin{figure*}[t!]
    \includegraphics[width=1\textwidth]{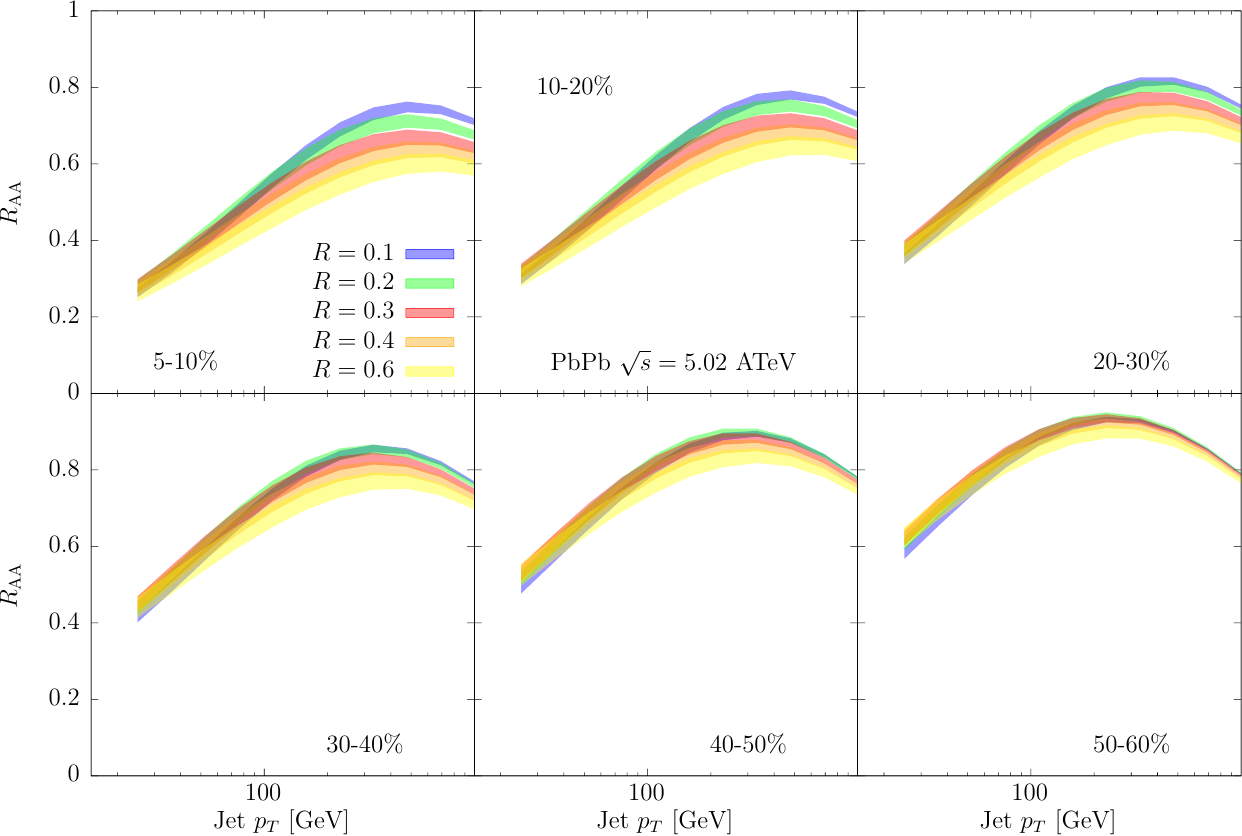}
     \includegraphics[width=1\textwidth]{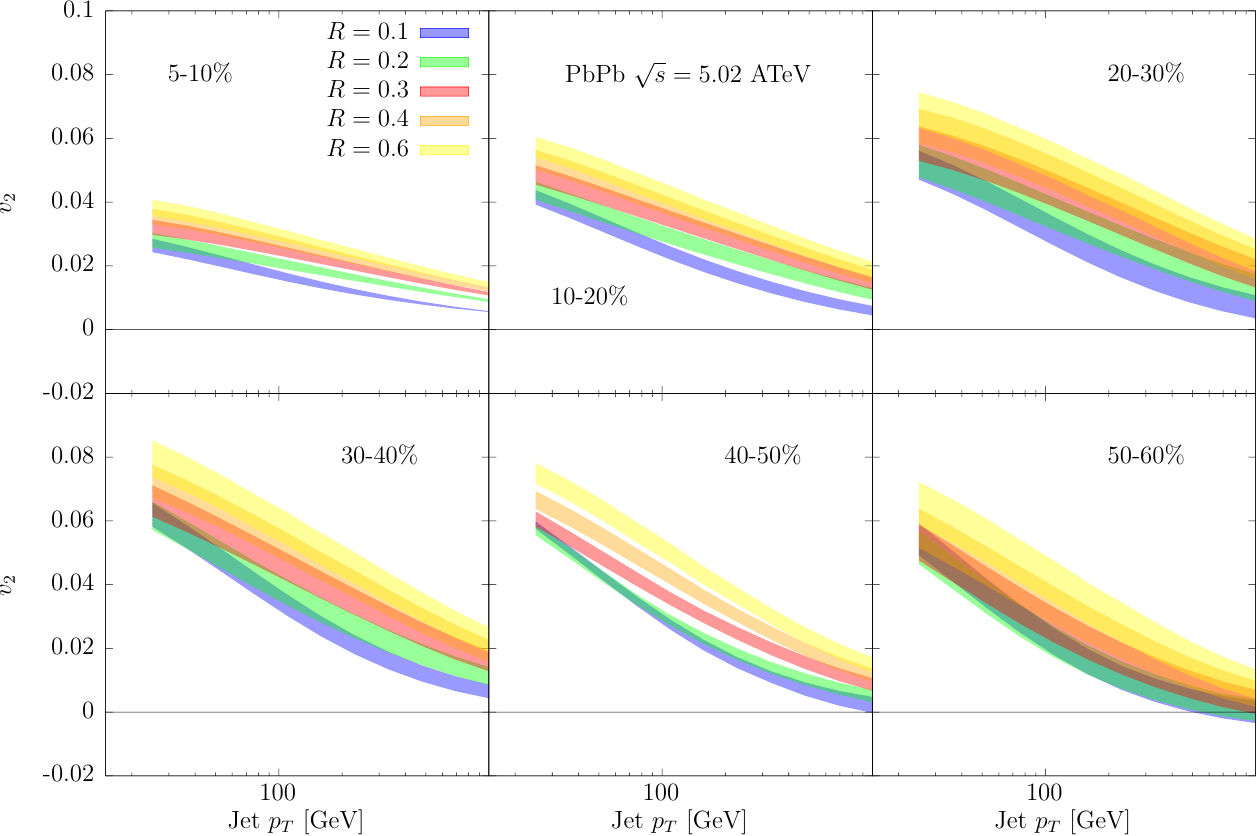}
    \caption{Jet $R_{\rm AA}(R,\pT)$ (top) and jet $v_2(R,\pT)$ (bottom) at LHC for PbPb collisions at $\sqrt{s}=5.02$ ATeV.}
    \label{fig:raalhc}
\end{figure*}

\begin{figure*}[t!]
    \includegraphics[width=1\textwidth]{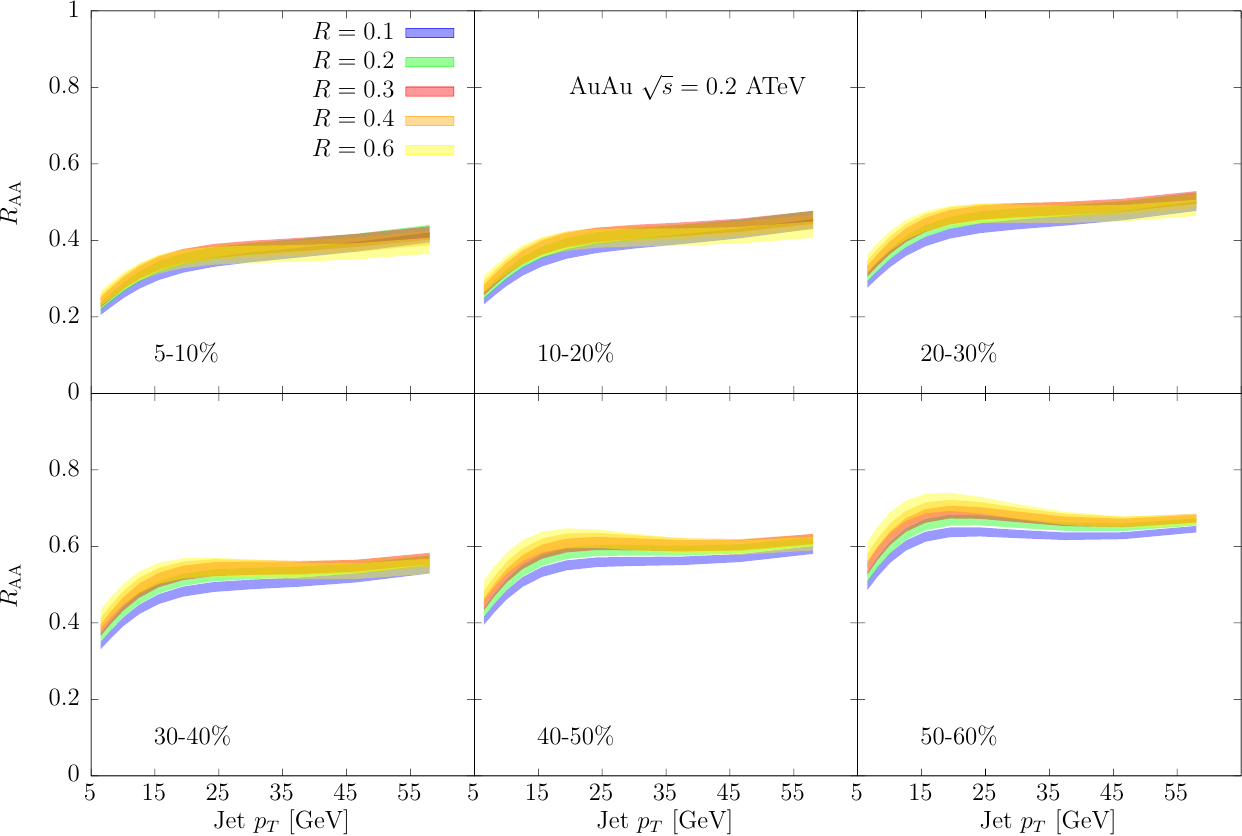}
     \includegraphics[width=1\textwidth]{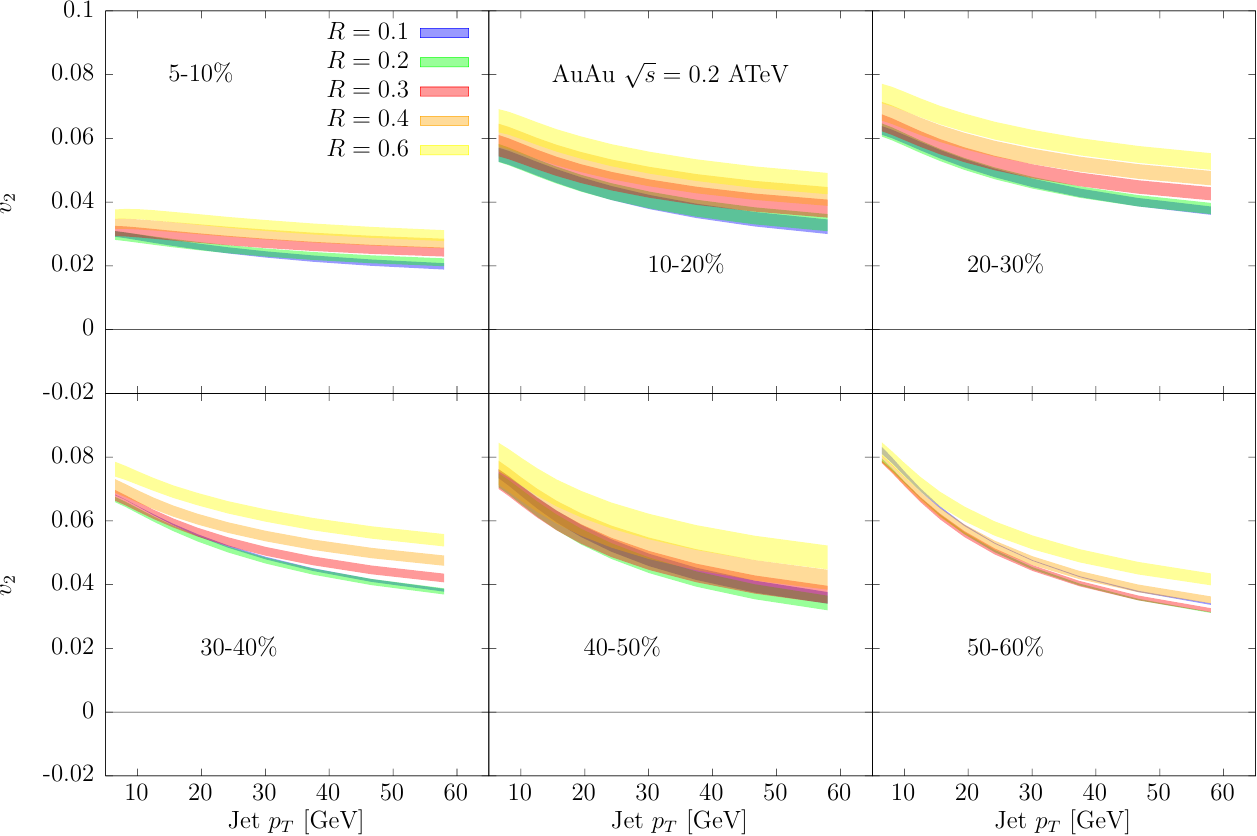}
    \caption{Jet $R_{\rm AA}(R,\pT)$ (top) and jet $v_2(R,\pT)$ (bottom) at RHIC for AuAu collisions at $\sqrt{s}=0.2$ ATeV.}
    \label{fig:raarhic}
\end{figure*}

\section{Full scan on the $R$ dependence of $R_{\rm AA}$ and jet $v_2$ with centrality at RHIC and the LHC.}
\label{sec:fullscan}
In this Section we show a comprehensive set of results for jet $R_{\rm AA}$ and jet $v_2$ for LHC (Fig.~\ref{fig:raalhc}) and RHIC (Fig.~\ref{fig:raarhic}) collision systems. In the following we comment on the main aspects of the $R$ and $\pT$ dependence of these observables, which offer complementary information to that presented in the main text.

We first discuss the LHC results shown in Fig.~\ref{fig:raalhc}. The results in the top panel indicate that jet $R_{\rm AA}$ for jets with different $R$ are much more similar at low $\pT$ than at high $\pT$, where larger $R$ jets are more suppressed, specially for the most central classes. This is because of the competition between phase-space and recapture of energy effects. The size of the resolved phase-space grows with jet $p_T$ and $R$. Even though a large $R$ jet is more capable of retaining energy emitted from its core, the larger amount of energy loss sources present at high-$\pT$ trump this effect and the net result is an increased suppression. In contrast, at low $\pT$ the recapture of the emitted energy plays a more important role, yielding overall a very mild dependence of suppression on $R$.
As we move towards more peripheral classes, the size of the resolved phase-space diminishes as the typical value of $\theta_c$ increases, yielding an increasingly flatter dependence with $R$ across a wider jet $\pT$ range. We even observe how for the most peripheral class, in the bottom right panel, large $R$ jets are somewhat less suppressed than small $R$ ones.

Interestingly, for the case of jet $v_2$ (lower panel of Fig.~\ref{fig:raalhc}) we in general find that $v_2(\pT,R')\geq v_2(\pT,R)$ if $R'>R$, for any centrality class. This is explained in detail with simple analytical estimates in Section~\ref{sec:discussion}, and is in essence related to the interplay between the marked length dependence of $\theta_c$ and the fact that jet $v_2$ is a length-differential jet suppression observable. As the typical value of $\theta_c$ increases with decreasing centrality, we find that jet $v_2$ values tend to be similar among jets with $R\leq \theta_c$. This is specially so towards lower jet $\pT\approx 50$ GeV, where phase-space effects are not too large. The important remark to make is that, in contrast to the jet $R_{\rm AA}$ results, relative differences in jet $v_2$ between small-$R$ and large-$R$ jets do in general remain large over a wide range of jet $\pT$ and centrality classes. 

By moving to the RHIC results of Fig.~\ref{fig:raarhic} we observe qualitatively similar trends. Jet $R_{\rm AA}$, shown in the top panel, is naturally covering a much narrower jet $\pT$ range due to smaller centre of mass energies. The relatively small size of the resolved phase-space implies that large-$R$ jets are slightly less suppressed than small-$R$ jets at fairly low values of jet $\pT$ (where the applicability of our framework should be put to question), while yielding a very mild dependence towards higher jet $\pT \approx 40$ GeV. Again, however, we find that jet $v_2$, shown in the lower panel of Fig.~\ref{fig:raarhic}, does present the same ordering discussed above for the LHC results. To appreciate the interesting decorrelation between jet $v_2$ and jet $R_{\rm AA}$ behavior with varying $R$, it is instructive to note that even though jet $R_{\rm AA}$ can increase (less quenching) with increasing $R$, jet $v_2$ will always increase with increasing $R$. The fact that the size of the resolved phase-space remains relatively small over the whole $\pT$ range makes the grouping of jets whose $R$ is smaller than the typical value of $\theta_c$ for a given centrality class more clear than it was the case for the LHC results of Fig.~\ref{fig:raalhc}.

We conclude by highlithing that the results of this Section imply that the $\theta_c^*/R$ scalings predicted in Section~\ref{sec:discussion} for jet $v_2$ ratios are a very transparent manifestation of the physics of color decoherence in jet suppression, dissimilar to the case of the jet $R_{\rm AA}$ ratios, where relative differences among jets with different $R$ are already small due to competing effects.

\end{document}

%% file: definitions.tex
\usepackage{dcolumn}
\usepackage{bm}
\usepackage{epstopdf}
\usepackage{mathrsfs}
\usepackage{amsmath,amssymb,amsfonts,latexsym}
\usepackage[normalem]{ulem}
\usepackage{amsmath,bbold}
\usepackage{xcolor}
\usepackage{array}
\usepackage{slashed}
\usepackage{nicefrac}
\usepackage[colorlinks=true,linktocpage=true,linkcolor=blue,citecolor=blue]{hyperref}
\usepackage[utf8]{inputenc}
\usepackage[mathcal]{euscript}

\def\del{\partial}

\def\sst{\scriptscriptstyle}
\newcommand{\eqn}[1]{Eq.~\eqref{#1}}

\long\def\comment#1{ }

\newcommand\redout{\bgroup\markoverwith{\textcolor{red}{\rule[.5ex]{2pt}{2pt}}}\ULon}

\def\0{{\boldsymbol 0}}

\def\and{\qquad\text{and}\qquad}

\def\gmed{g_{\rm med}}

\def\tform{{t_\text{f}}}
\def\tdecoh{t_\text{d}}

\def\jet{\text{jet}}

\def\min{\text{min}}
\def\max{{\text{max}}}

\def\Qbare{Q^{(0)}}
\def\pT{p_{\sst T}}

\newcommand{\beq}{\begin{eqnarray}}
\newcommand{\eeq}{\end{eqnarray}}
\newcommand{\be}{\begin{eqnarray*}}
\newcommand{\ee}{\end{eqnarray*}}
\newcommand{\bal}{\begin{align}}
\newcommand{\eal}{\end{align}}
\newcommand{\dd}{{\rm d}}
\newcommand{\rme}{{\rm e}}

\def\rmR{{\rm Re}}

\def\abar{{\rm \bar\alpha}}

\def\PS{\Omega_{\rm res}}
\def\PSincoh{\Omega_{\rm incoh}}

\newcommand{\nn}{\nonumber\\ }

%% file: main.bbl
%